\documentclass[sn-mathphys-num]{sn-jnl}
\usepackage{graphicx}%
\usepackage{multirow}%
\usepackage{multicol}
\usepackage{amsmath,amssymb,amsfonts}%
\usepackage{amsthm}%
\usepackage{mathrsfs}%
\usepackage[title]{appendix}%
\usepackage{enumitem}
\usepackage{subcaption}
\usepackage{float}
\usepackage{siunitx}
\usepackage{times,xcolor,stackengine}
\usepackage{pgfplots}
\usetikzlibrary{backgrounds}
\usetikzlibrary{decorations.text,decorations.markings,decorations.pathmorphing,decorations.pathreplacing}
\usetikzlibrary {arrows.meta} 
\definecolor{darkorange}{rgb}{1.0, 0.55, 0.0}
\definecolor{newred}{RGB}{125,0,45}
\definecolor{newblue}{RGB}{0,80,158}
\definecolor{newgray}{gray}{0.4}
\definecolor{ArmyGreen}{rgb}{0.29, 0.33, 0.13}
\definecolor{BostonRed}{rgb}{0.9, 0.0, 0.0}
\definecolor{CadmiumGreen}{rgb}{0.0, 0.42, 0.24}
\definecolor{darkyellow}{RGB}{205,205,0}
\definecolor{Gray}{gray}{0.9}
\definecolor{LightCyan}{rgb}{0.88,1,1}
\definecolor{PastelGreen}{rgb}{0.47,0.87,0.47}
\definecolor{PastelRed}{rgb}{0.87,0.47,0.47}
\definecolor{PastelYellow}{rgb}{0.99, 0.99, 0.59}
\definecolor{platinum}{rgb}{0.9, 0.89, 0.89}
\definecolor{pistachio}{rgb}{0.58, 0.77, 0.45}
\definecolor{princetonorange}{rgb}{1.0, 0.56, 0.0}
\definecolor{FigureBlue}{rgb}{0.447, 0.624, 0.812}
\definecolor{NewBlue}{rgb}{0.712, 0.806, 0.861}
\definecolor{LightBlue}{rgb}{0.812, 0.906, 0.961}
\definecolor{airforceblue}{rgb}{0.36, 0.54, 0.66}
\definecolor{amber}{rgb}{1.0, 0.75, 0.0}
\usepackage{url} 
\colorlet{darkred}{red!85!black}
\colorlet{darkgreen}{green!50!black}
\colorlet{darkblue}{blue!60!black}
\hypersetup{
colorlinks   = true, 
urlcolor     = darkgreen, 
linkcolor    = darkblue, 
citecolor   = darkred 
}
\usepackage{titlesec} 
\usepackage{bm,dsfont}
\usepackage[scr=boondoxo,scrscaled=1.05]{mathalfa}
\usepackage{placeins}
\usepackage{epstopdf}
\usepackage{siunitx}

\theoremstyle{thmstyleone}%
\theoremstyle{thmstyletwo}%

\theoremstyle{thmstylethree}%
\newtheorem{definition}{Definition}%

\numberwithin{equation}{section}

\titleformat{\subsubsection}[runin]
  {\normalfont\normalsize\bfseries}{\thesubsubsection}{1em}{}

\begin{document}

\title[Decay of the survival probability in realistic multi-qubit platforms]{Decay of the survival probability of a local excitation in  multi-qubit platforms}


\author*[1]{\fnm{Paolo} \sur{Muratore-Ginanneschi}}\email{paolo.muratore-ginanneschi@helsinki.fi}

\author[2,3]{\fnm{Bayan} \sur{Karimi}}\email{karimib@uchicago.edu}

\author[3]{\fnm{Jukka} \sur{Pekola}} \email{jukka.pekola@aalto.fi}

\affil[1]{\orgdiv{Department of Mathematics and Statistics}, \orgname{University of Helsinki}, \orgaddress{\city{Helsinki}, \postcode{P.O. Box 68, 00014}, \country{Finland}}}

\affil*[2]{\orgdiv{Pritzker School of Molecular Engineering}, \orgname{University of Chicago}, \orgaddress{ \city{Chicago}, \postcode{IL 60637}, \country{USA}}}

\affil*[3]{\orgdiv{Pico group, QTF Centre of Excellence, Department of Applied Physics}, \orgname{Aalto University School of Science}, \orgaddress{ \city{Espoo}, \postcode{P.O. Box 13500, 00076 Aalto}, \country{Finland}}}


\abstract{We present a theoretical study of the survival probability of a state initially prepared in the one-particle sector of a multi-qubit system. The motivation for our work is the ongoing laboratory development of multi-qubit platforms based on superconducting circuits. 
Using elementary concepts of random matrix theory, we obtain analytic expressions for the survival probability in mathematical models of platforms which, albeit stylized, have been previously shown to provide relevant benchmarks for experimental data. In particular, we show that the decay properties are sensitive to the property of the Hamilton operator to have extended states. The survival probability does not appear instead to depend on whether the interaction between qubits is described by a Gaussian orthogonal ensemble (often interpreted as a model of ''chaotic'' dynamics) or is modeled by an analytically solvable chain. We interpret this phenomenon as a manifestation of a general mechanism for the emergence of equilibration in purely unitary dynamics. Finally, under the same hypothesis of an initial preparation with projection on a large fraction of the extended eigenstates of the Hamilton operator, we show how to extend the classical Kac-Mazur-Montroll estimate of the return time to the quantum survival probability.     
  }




\maketitle

\section{Introduction}
\label{sec:intro}

At present, superconducting qubits are, probably, the most promising platform to construct quantum information processing devices and fault tolerant quantum computers \cite{ClWi2008,DeSc2013,KjScBrKrOl2020}. Contemporary error-protection architectures are based on assemblies of physical qubits that together form logical qubits \cite{CleA2022}. These assemblies involve a large number of qubits: in 2024 \cite{BrCrGaMaRaYo2024} demonstrated that a full error correction protocol for 12 logical qubits during nearly 1 million syndrome cycles requires 288 physical qubits in total. The technological feasibility of building large assemblies of qubits also paves the way for exploring questions that are both relevant to quantum computing and of fundamental interest to quantum theory \cite{ClWi2008}.  

From the mathematical point of view, the qualitative properties of the dynamics of multi-qubit systems  as those of any quantum system with pure point spectrum \cite{BoLo1957,SchL1978} are determined by the theory of almost periodic functions, initiated by Harald Bohr \cite{BohH1925} in the '20s of last century. 
 We refer to \cite{BesA1954} for a classical introduction or to \cite{CorC2009} for a contemporary exposition of the theory of almost periodic functions.  Roughly speaking, almost periodic functions are sums over trigonometric functions and, when the sum extends to infinite addends, absolutely converging Fourier series.  The properties of almost periodic functions are central to understanding the emergence of irreversibility in unitary and conservative dynamics, or, in the words of Robert Zwanzig \cite[\S~10]{ZwaR2001} the {\textquotedblleft}paradoxes of irreversibility{\textquotedblright}. Famously, Zermelo \cite{ZerE1896} objected -- the so called {\textquotedblleft}Wiederkehreinwand{\textquotedblright} (i.e. recurrence objection)-- to Boltzmann's derivation of the second law of thermodynamics by pointing out how the idea of relaxation to equilibrium in strictly mathematical terms contradicts Poincar\'e’s recurrence theorem. The recurrence theorem stipulates that the dynamics of an isolated Hamiltonian system starting from any point of a bounded phase space will eventually return arbitrarily close to the initial condition, see, e.g., \cite[\S~8.5]{FaMa2006}. Even more harshly,  in his treatise on thermodynamics \cite{PoiH1892_2}, Poincar\'e  deliberately ignored Boltzmann's statistical explanation of the second law considering it untenable within mechanistic hypotheses. Boltzmann's derivation of the second law can be reconciled with Poincar\'e's theorem by considering  that the recurrence time grows exponentially with the number of frequencies entering an almost periodic function. In his reply to Zermelo \cite{BolL1896}, Boltzmann recognized the role of this phenomenon. Rigorous mathematical results supporting Boltzmann's reply came half a century later from the works \cite{KacM1943,KacM1947} where Kac applies the tools provided by the theories of almost periodic functions and stochastic processes introduced in the meantime. A detailed historical reconstruction of the XIX debate is available in \cite{vStM2013} whereas a thorough discussion of the implications for the foundations of statistical mechanics, classical and quantum, can be found in \cite{BaGrVuZa2025}. In the  second half of the XX century, apparent relaxation to an {\textquotedblleft}equilibrium{\textquotedblright} in between revivals of almost periodic dynamics and Kac's estimates of the recurrence time have been applied in important contributions, in particular \cite{DaLoPr1962,ZurW1982}, aimed at reconciling quantum measurement process with unitary dynamics, see however \cite{BuLaMi1996,BaGh2000}. In particular, \cite{ZurW1982} has reached  a greater consensus as a model to explain the emergence of macroscopic phenomena from microscopic quantum dynamics in a given environment \cite[\S~15]{GhiG2007}. 
  
  
Recently \cite{KaWuClPe2026}, two of us with co-authors designed an experiment to test theoretical predictions for the survival probability of a state of a central qubit coupled to an ensemble of $N$ {\textquotedblleft}environmental{\textquotedblright} qubits. In particular, \cite{KaWuClPe2026} discusses the central qubit dynamics in mathematical models adapted to describe  the expected phenomenology in recently introduced tunable experimental platforms \cite{WuYaAnAnCle2024,AnAbMi2025}, which allow for varying energy partitioning and coupling strengths between superconducting qubits.
 The motivation and results of \cite{KaWuClPe2026} have many points of contact with the study of the survival probability of a central Fermion under calorimetric measurement slightly earlier co-authored by the third of us \cite{DoGoMG2020}.  It is fair to add, that there exist many classical studies of the survival probability of resonances see e.g. \cite{FoGhRi1978,PerA1980,GaVi1988} and especially targeted at identifying time scales where genuine quantum behavior can be observed beyond the exponential decay intermediate asymptotics, see e.g. \cite{AnaC2018}. Nevertheless, predictions adapted to realistic experimental platforms maintain a fundamental interest. This is because measuring the survival probability of a pure state of a many body system over long time scales constitutes a test of quantum mechanics that still poses considerable experimental challenges. To-date, very few experimental results have been documented, e.g. \cite{RoHiMo2006,CrPeFaSc2019}.
 
 In this contribution, we refine the analytical investigation of the single particle sector in the same mathematical model of multi-qubit dynamics considered in \cite{KaWuClPe2026}. In particular, we study the survival probability of a local excitation: the probability of finding a reference qubit, called the central qubit, in the excited state at any time later than the initial when the central qubit is prepared in the excited state while all the surrounding qubits are in the ground state. Despite its simplicity, this preparation already gives rise to a non-trivial phenomenology depending on the strength and geometry of the interactions. We obtain three main results, which, although relatively straightforward, to the best of our knowledge, have not been previously reported in the literature. To start with, parameters such as the energy splitting and coupling constants between qubits fluctuate mostly because of limits in fabrication accuracy. Mathematically, this translates into modeling these quantities as random variables with finite variance. Under this hypothesis, we show that if the one-particle sector is described by a structured random matrix, with independent entries but with diagonal elements obeying a different distribution than off-diagonal ones, the survival amplitude does not fluctuate in the limit of infinite multi-qubit system. This is expected from a thermodynamic limit point of view. More interestingly, in the same limit, we prove that  if the univariate eigenvalue distribution of the environment is uniform, the survival probability is exactly that of the non-relativistic Lee model \cite{LeeT1954} and \cite{JaKrPi2006,WolT2013}.  This is a very useful result for experimental applications because it provides a benchmark to understand dynamical regimes that are not accessible with perturbative arguments. The second result is that the survival probability for the dynamics generated by a Gaussian orthogonal ensemble with a deterministic shift on the diagonal is exactly the same as that of an analytically solvable qubit chain with all equal first neighbors interactions in the limit of infinite multi-qubit system. These models share the property of admitting only extended normal modes, all with a non-vanishing projection on the localized initial state of the system. Hence, our result is a further illustration of the fact that equipartition depends on the property of an observable to excite delocalized normal modes of the Hamiltonian generating the dynamics and not on the statistics of the eigenvalue level spacings of the Hamiltonian\cite{CaBaLuMGVu2025}. The third result goes in the same direction. Namely, we extend Kac's estimate  of the return time \cite{KacM1943} to the survival probability of a local excitation of a quantum system with $N$-states. The crucial assumption is again the presence of delocalized eigenstates with entries of order $O(N^{-1/2})$. This hypothesis is satisfied by a large class of random matrix models, including those of direct interest for us \cite{RuVe2015}. We extend Kac's estimate by adapting to the quantum case the approach followed by Mazur and Montroll in \cite{MaMo1960}. We refer to \cite{KoOs2026} for an alternative derivation of an upper bound on the return time of the survival probability of an initial preparation of a quantum system with a finite number of states.
 
 The structure of the paper is as follows. In section~\ref{sec:Fock} we recall the Fock space representation of a multi-qubit system. The peculiarity of qubits is that they obey a {\textquotedblleft}parafermion statistics{\textquotedblright} \cite{WuLi2002} that mixes the properties of fermions and bosons. In section~\ref{sec:Fock} we also introduce the models of our main interest. In section~\ref{sec:sp} we introduce the survival probability of a local excitation and recall its general properties. In section~\ref{sec:sp-explicit} we study the survival probability of a fully analytically solvable chain model, which, although very stylized, constitutes a useful benchmark for experimental data. This is instructive because, in the limit of infinite sized one-particle sector, the exact expression tends to a Bessel function. This is a concrete illustration of how dissipative behavior emerges when we push to infinity the return time of an almost periodic motion. In particular, it shows that before the first revival the predictions obtained in the infinite system limit are very accurate also for relatively small values of $N$. In section~\ref{sec:mqm}, we start by recalling a general method to compute survival probabilities in quantum mechanics. Next, we briefly illustrate how to apply this method to systematically extend the perturbative analysis of \cite{KaWuClPe2026}. Finally, we turn to the derivation of our first two main results. In section~\ref{sec:Zeno} we show how the lower bound on the survival probability provided by the Mandelstam-Tamm inequality captures the decay of the survival probability for models where the van Hove exponential decay intermediate asymptotics is absent. We present our third main result, the extension of the Kac-Mazur-Montroll estimate of the revival time to the quantum survival probability,  in section~\ref{sec:Kac}. We add some complementary information used to derive the main results in the appendices. The last section is devoted to conclusions and outlook.

\section{The Fock space of qubits}
\label{sec:Fock}

As for fermions and bosons, the occupation number representation provides a complete orthonormal basis spanning the Hilbert space of a system of $N$ qubits, i.e. identical particles with two levels. We construct this basis by acting on the vacuum with creation and annihilation operators defined by taking tensor products of corresponding single particle operators. To fix our conventions, we identify the canonical basis of  $\mathbb{C}^2$
\begin{align}
&	e_{1}=\begin{bmatrix}
	1	\\ 0  
	\end{bmatrix}
&&
e_{2}=\begin{bmatrix}
	0	\\ 1  
\end{bmatrix}
\end{align}
with the eigenstates of the single particle Hamilton operator
\begin{align}
	\operatorname{H}=\epsilon_{1}\sigma_{+}\sigma_{-}
	=\begin{bmatrix}
	\epsilon_{1} & 0	\\ 0 & 0  
	\end{bmatrix}
\end{align}
Here and throughout the text, we set $\hbar$ to one and we define
\begin{align}
&	\sigma_{+}=\begin{bmatrix}
	0	& 1\\ 0 & 0 
	\end{bmatrix}
	&&
	\sigma_{-}=\begin{bmatrix}
		0	& 0\\ 1 & 0 
	\end{bmatrix}
\end{align}
In such a case, the vacuum state of the $N$ particle system is
\begin{align}
	\Phi_{0}=\underbrace{e_{2}\otimes e_{2}\otimes\dots \otimes\,e_{2}}_{N-\mbox{times}}
\end{align}
Correspondingly, we construct creation and annihilation operators of the $k$-th qubit by taking tensor products of operators on the individual particle Hilbert spaces
\begin{subequations}
	\label{Fock:ladder}
	\begin{align}
		&\label{Fock:low}
		\operatorname{a}_{k}=\underbrace{\operatorname{1}_{2}\,\otimes\,\dots\operatorname{1}_{2}}_{k-1}\,\otimes\,\sigma_{-}\,\otimes\,\underbrace{\operatorname{1}_{2}\,\otimes\,\operatorname{1}_{2}}_{N-k}
		\\
		&\label{Fock:up}
		\operatorname{a}_{k}^{\dagger}=\underbrace{\operatorname{1}_{2}\,\otimes\,\dots\operatorname{1}_{2}}_{k-1}\,\otimes\,\sigma_{+}\,\otimes\,\underbrace{\operatorname{1}_{2}\,\otimes\,\operatorname{1}_{2}}_{N-k}
	\end{align}
\end{subequations}
 Using these operators, we derive the occupation number representation of the multi-qubit Hilbert space as
\begin{align}
	\Phi_{\bm{n}}=\prod_{i=1}^{N}\operatorname{a}_{i}^{\dagger}{}^{n_{i}}\Phi_{0}
	\label{Fock:basis}
\end{align}
The collection of $\left\{ n_{i} \right\}_{1}^{N}$ in (\ref{Fock:basis}) forms a $N$-tuple $\bm{n}$  whose elements only take $0$ or $1$ values:
\begin{align}
&	\bm{n}=[n_{1},\dots,n_{N}] && n_{i}\in\left\{ 0,1 \right\}\hspace{0.5cm}\forall\,i=1,\dots,N
\end{align}
We may regard the $2^{N}$ states (\ref{Fock:basis}) as eigenstates of the non-interacting Hamiltonian
\begin{align}
	\operatorname{H}=\sum_{i=1}^{N}\epsilon_{i}\operatorname{a}_{i}^{\dagger}\operatorname{a}_{i}
	\label{Fock:Hni}
\end{align} 
for any choice of the $\epsilon_{i}$'s. In addition, symmetrization of  (\ref{Fock:basis}) under particle exchange produces a Dicke state vector. Efficient preparation of Dicke states using quantum circuits of finite depth and local operations and classical communication (LOCC) is an active field of research \cite{WaTe2021,BuFoLoNe2024,PiStCi2024}.   

So far the occupation number representation of a multi-qubit Hilbert space appears similar to that of fermions. A profound difference appears at the level of the commutation relations:
\begin{align}
	&	\left[\operatorname{a}_{\ell}\,,\operatorname{a}_{k}\right]=\left[\operatorname{a}_{\ell}\,,\operatorname{a}_{k}^{\dagger}\right]=0
	&& \forall \ell \neq k
	\label{Fock:pf1}
\end{align}
but
\begin{align}
\begin{split}
&	\left[\operatorname{a}_{\ell}\,,\operatorname{a}_{\ell}^{\dagger}\right]_{+}=\operatorname{1}_{2^{N}}
\\
&\operatorname{a}_{\ell}^{2}=\operatorname{a}_{\ell}^{\dagger}{}^{2}=0
\end{split}    
&&\forall \ell=1,\dots,N
	\label{Fock:pf2}
\end{align}
where $\left[\cdot\,,\cdot\right]_{+}$ denotes the anticommutator operation. In words, ladder operators on different single-particle Hilbert spaces commute like Boson ladder operators, but those acting on the same single-particle Hilbert space anticommute like Fermion ladder operators. For this reason, the commutation relation (\ref{Fock:pf1}), (\ref{Fock:pf2}) is sometimes referred to as  {\textquotedblleft}parafermionic{\textquotedblright} \cite{WuLi2002}.
The consequence of parafermionic commutation relations becomes manifest when considering interacting qubits.
The simplest situation occurs if we consider general normal ordered time independent quadratic interactions. In such a case, the Hamilton operator reads
\begin{align}
	\operatorname{H}=\sum_{i,j=1}^{N}\operatorname{a}_{i}^{\dagger}\operatorname{H}_{i,j}^{\scriptscriptstyle{(BG)}}\operatorname{a}_{i}
	\label{Fock:qH}
\end{align}
where $\operatorname{H}$ is an $N\,\times\,N$ self-adjoint matrix, which in the condensed matter literature is referred to as the Bogolyubov-de Gennes's matrix \cite{LeMaMuGr2020}. It is straightforward to verify that the Hamilton operator (\ref{Fock:qH}) commutes with the number operator
\begin{align}
	\operatorname{N}=\sum_{i=1}^{N}\operatorname{a}_{i}^{\dagger}\operatorname{a}_{i}
\end{align}
for any choice of the Bogolyubov-de Gennes's matrix:
\begin{align}
	\left[\operatorname{H}\,,\operatorname{N}\right]=0 
	\label{Fock:quanta}
\end{align}
This is because Pauli matrices provide a representation of the  $\mathfrak{su}(2)$ Lie algebra
\begin{align}
	\left[\operatorname{a}_{k}^{\dagger}\operatorname{a}_{k}\,,\operatorname{a}_{k}^{\dagger}\right]=\operatorname{a}_{k}^{\dagger}
\end{align}
For boson and fermions, the straightforward way to obtain the spectral decomposition of a quadratic Hamilton operator is to construct the single (composite-)particle sector of the theory by diagonalizing the Bogolyubov-de Gennes matrix. Specifically, one solves the eigenvalue problem
\begin{align}
&	\operatorname{H}^{\scriptscriptstyle{(BG)}}\, v_{i}=h_{i}\,v_{i} && i=1,\dots,N
\end{align} 
in order to reduce (\ref{Fock:qH}) to a non-interacting Hamiltonian in terms of composite particles created or annihilated by ladder operators specified by the Bogolyubov-Valatin transformation \cite[\S~15.9]{DiCaRa2015}. By following this procedure for qubits, we arrive at
\begin{align}
	\operatorname{b}_{\ell}=\sum_{i=1}^{N}v_{\ell,i}\operatorname{a}_{i}
	\nonumber
\end{align}
where the sum ranges over the components of the eigenvectors of the Bogolyubov de Gennes Hamiltonian. We obtain
\begin{align}
	\operatorname{H}=\sum_{\ell=1}^{N}h_{\ell}\operatorname{b}_{\ell}^{\dagger}\operatorname{b}_{\ell}
\end{align}
Unfortunately, the simplification is only apparent as the ladder operators do not obey any definite commutation relation, in consequence of the fact that qubits are neither bosons nor fermions:
\begin{align}
	\left[\operatorname{b}_{\ell}\,,\operatorname{b}_{k}^{\dagger}\right]=\delta_{\ell,k}-2\sum_{i,j=1}^{N}v_{\ell,i}\bar{v}_{k,i}\operatorname{a}_{i}^{\dagger}\operatorname{a}_{i}
\end{align}
An approximate bosonic behavior emerges if we evaluate the commutator on the first excited states of large systems under the hypothesis that the Bogolyubov de Gennes Hamiltonian has only extended eigenstates:
\begin{align}
	v_{\ell,i}\sim O(N^{-1/2})
\end{align}
Indeed, the protocol \cite{PiStCi2024} to prepare  eigenstates of spin models using finite-depth circuits and LOCC relies on this approximation.

In general, the Bogolyubov-Valatin transformation does not produce the exact diagonalization of the many-qubit Hamilton operator (\ref{Fock:qH}). A notable exception occurs when (\ref{Fock:qH}) only includes nearest neighbor interactions, i.e., in the case of qubit chains. In the latter case, the Jordan-Wigner transform maps a qubit chain into a Fermion chain. To verify this claim, we observe that $N$-fermion ladder operators admit the tensor product representation
\begin{subequations}
	\label{Fock:Fermi}
	\begin{align}
		&\label{Fock:Fermi1}
		\operatorname{c}_{k}=\underbrace{\sigma_{3}\,\otimes\,\dots\sigma_{3}}_{k-1}\,\otimes\,\sigma_{-}\,\otimes\,\underbrace{\operatorname{1}_{2}\,\otimes\,\operatorname{1}_{2}}_{N-k}
		\\
		&\label{Fock:Fermi2}
		\operatorname{c}_{k}^{\dagger}=\underbrace{\sigma_{3}\,\otimes\,\dots\sigma_{3}}_{k-1}\,\otimes\,\sigma_{+}\,\otimes\,\underbrace{\operatorname{1}_{2}\,\otimes\,\operatorname{1}_{2}}_{N-k}
	\end{align}
\end{subequations}
where the $\operatorname{c}_{k}$ and $\operatorname{c}_{k}^{\dagger}$ satisfy the canonical anticommutation relations.
The immediate consequences are that
\begin{align}
	\operatorname{c}_{k}^{\dagger}\operatorname{c}_{k}=\operatorname{a}_{k}^{\dagger}\operatorname{a}_{k}
\end{align}
and
\begin{align}
	\operatorname{a}_{k}^{\dagger}=
	\prod_{j<k}\left(2\,\operatorname{c}_{j}^{\dagger}\operatorname{c}_{j}-\operatorname{1}_{2^{N}}\right)\,\operatorname{c}_{k}^{\dagger}=\operatorname{c}_{k}^{\dagger}\prod_{j<k}\left(2\,\operatorname{c}_{j}^{\dagger}\operatorname{c}_{j}-\operatorname{1}_{2^{N}}\right)
\end{align}
The Wigner-Jordan transformation is a non-local involution:
\begin{align}
	\operatorname{c}_{k}^{\dagger}=\operatorname{a}_{k}^{\dagger}\prod_{j<k}\left(2\,\operatorname{a}_{j}^{\dagger}\operatorname{a}_{j}-\operatorname{1}_{2^{N}}\right)
\end{align}
The Jordan-Wigner transform allows us to extend the exact results obtained for Kitaev chains in \cite{LeMaMuGr2020} to qubits. In particular, it allows us to test the predictions of \cite{DoGoMG2020} by implementing a chain interaction geometry on the multi-qubit platforms \cite{WuYaAnAnCle2024,AnAbMi2025}. 

We refer to \cite[\S~8.2]{ShaR2017} and \cite[\S~10]{SacS2011} for further details and applications of the Jordan-Wigner transform. 

\subsection{Example: qubit chain}
\label{sec:Fock-exe1}

For a large class of Bogolyubov - de Gennes matrices describing nearest-neighbor interactions, it is possible to write explicit expressions of eigenvalues and eigenvectors \cite{LosL1992}. A reference example is the qubit chain
\begin{align}
	\operatorname{H}=\omega\,\sum_{i=1}^{N} \operatorname{a}_{i}^{\dagger}\operatorname{a}_{i}+g\sum_{i=1}^{N-1}
	\Big{(}\operatorname{a}_{i+1}^{\dagger}\operatorname{a}_{i}+\operatorname{a}_{i}^{\dagger}\operatorname{a}_{i+1}\Big{)}
	\label{Fock-exe1:H}
\end{align}
The structure of the interactions is preserved by the Jordan-Wigner transform
\begin{align}
	\operatorname{H}=\sum_{i=1}^{N} \omega\,\operatorname{c}_{i}^{\dagger}\operatorname{c}_{i}-g\sum_{i=1}^{N-1}
	\Big{(}\operatorname{c}_{i+1}^{\dagger}\operatorname{c}_{i}+\operatorname{c}_{i}^{\dagger}\operatorname{c}_{i+1}\Big{)}
\end{align}
This fact allows us to reduce the spectral problem to that of the Bogolyubov de Gennes matrix whose eigenvalues are
\begin{align}
&	h_{\ell}=\omega+2\,g\,\cos\frac{\ell\, \pi}{N+1} && \ell=1,\dots,N
	\label{Fock-exe1:ev}
\end{align}
The corresponding normalized eigenvectors are
\begin{align}
	& v_{\ell,k}=\frac{1}{\sqrt{Z_{\ell}}}\sin \frac{\ell\, k\,\pi}{N+1}&& \ell,k=1,\dots,N
		\label{Fock-exe1:es}
\end{align}
A straightforward calculation proves that
\begin{align}
&	Z_{\ell}=Z:=\frac{N+1}{2} && \forall\,\ell=1,\dots,N
\end{align}
Composite particle ladder operators take the form
\begin{align}
	\operatorname{c}_{\ell}=\sum_{k=1}^{N}v_{\ell,k}\operatorname{a}_{k}^{\dagger}\prod_{j<k}\left(2\,\operatorname{a}_{j}^{\dagger}\operatorname{a}_{j}-\operatorname{1}_{2^{N}}\right)
\end{align}
and obey canonical anticommutation relations.

In section~\ref{sec:sp-explicit} we use the above elementary results to illustrate the time evolution of the survival probability of a local excitation of a multi-qubit system over all time scales. 

\subsection{An experimentally realistic model}
\label{sec:Fock-exe2}

The mathematical model of the multi-qubit experimental platform considered  \cite{KaWuClPe2026} is described by the Hamilton operator
\begin{align}
	\operatorname{H}=\sum_{i=1}^{N}\epsilon_{i}\operatorname{a}_{i}^{\dagger}\operatorname{a}_{i}+\sum_{i=1}^{N}\sum_{ j>i}^{N}
	g_{i,j}(\operatorname{a}_{i}^{\dagger}\operatorname{a}_{j}+\operatorname{a}_{j}^{\dagger}\operatorname{a}_{i})
	\label{Fock-exe2:H}
\end{align}
The $\epsilon_{i}$'s denote the qubits' energy splittings and the coupling constants $g_{i,j}$'s are real valued. 
The Hamilton operator (\ref{Fock-exe2:H}) allows all qubits to interact with each other, in principle, with different intensities. The qubit chain (\ref{Fock-exe1:H}) is thus a particular case. We refer to \cite{WuYaAnAnCle2024} for the derivation
of (\ref{Fock-exe2:H}) from circuit quantization and using the Schrieffer-Wolff transformation, see e.g. \cite{BrDiViLo2011},  to determine the effective couplings between the qubits energy modes.  Further approximations introduced in \cite{KaWuClPe2026} are the rotating wave approximation and neglecting anharmonic terms as well as any time dependence in the interaction.

If we project the Hamiltonian on the single qubit occupation basis (\ref{Fock:basis}) it reduces to a $2^{N}\,\times\,2^{N} $ block diagonal self-adjoint matrix. This is a consequence of the conservation of the number of quanta implied by (\ref{Fock:quanta}):
\begin{equation}
	\operatorname{H}=	
		\left[
		\begin{array}{ccccccc}
				\sum_{i=1}^{N}\epsilon_{i} &  0 & 0 & \dots   & \dots & \dots & 0 \\
				0 & \operatorname{H}_{\scriptscriptstyle{N-1}}^{\scriptscriptstyle{(N)}} & 0 & \dots & \dots & \dots & 0 \\
				0 &\hspace{-0.5cm} 0& \operatorname{H}_{\scriptscriptstyle{N-2}}^{\scriptscriptstyle{(N)}}&0 & \dots &  \dots&  0  \\
				 \vdots& \hspace{-0.5cm}\ddots &\hspace{-0.7cm}\ddots  & \hspace{-0.5cm}\ddots & \hspace{-0.5cm}\ddots & \hspace{-0.5cm}\ddots  &  \vdots \\
				 \vdots&  \hspace{-0.5cm}\ddots & \hspace{-0.5cm}\ddots&\hspace{-0.5cm}0 &\operatorname{H}_{\scriptscriptstyle{2}}^{\scriptscriptstyle{(N)}}& \hspace{-0.3cm}0 & 0\\
				0 & \hspace{-0.5cm}\ddots & \hspace{-0.5cm}\ddots   &\hspace{-0.5cm}\ddots  & \hspace{-0.3cm}0&  \operatorname{H}_{\scriptscriptstyle{1}}^{\scriptscriptstyle{(N)}}& 0 \\
				0 & \dots &  \dots  & \dots &  \hspace{-0.3cm}0&  \hspace{-0.3cm}0 & 0 \\
			\end{array}
		\right]
\end{equation} 
The blocks $\operatorname{H}_{\scriptscriptstyle{k}}^{\scriptscriptstyle{(N)}}$ on the diagonal are $d_{\scriptscriptstyle{k}}^{\scriptscriptstyle{(N)}}\,\times\,d_{\scriptscriptstyle{k}}^{\scriptscriptstyle{(N)}}$ square matrices whose row length is
\begin{align}
d_{\scriptscriptstyle{k}}^{\scriptscriptstyle{(N)}}=\frac{N!}{k! \,(N-k)!}
\end{align}
The diagonal entries in each of the blocks $ \operatorname{H}_{\scriptscriptstyle{k}}^{\scriptscriptstyle{(N)}}$ are the sum of the energy level separations (splittings) of the excited qubits. The non-vanishing off-diagonal entries describe interactions. For instance, in the case  $N=4$ the Hamilton operators of the non-trivial blocks are 
\begin{align}
&	\operatorname{H}_{\scriptscriptstyle{1}}^{\scriptscriptstyle{(4)}}=\begin{bmatrix}
		 \epsilon_{4} & g_{3,4} & g_{2,4} &  g_{1,4}  \\
		 g_{3,4} & \epsilon_{3}& g_{2,3} & g_{1,3}    \\
	     g_{2,4} & g_{2,3} & \epsilon_{2} & g_{1,2}  \\
		 g_{1,4} & g_{1,3} & g_{1,2} & \epsilon_{1} 
	\end{bmatrix}
&&	\operatorname{H}_{\scriptscriptstyle{3}}^{\scriptscriptstyle{(4)}}=\begin{bmatrix}
	\epsilon_{2} +\epsilon_{3} +\epsilon_{4} & g_{1,2} & g_{1,3} &  g_{1,4}  \\
	g_{1,2} & \epsilon_{1} +\epsilon_{3}+\epsilon_{4} & g_{2,3} & g_{2,4}    \\
	g_{1,3} & g_{2,3} & \epsilon_{1} +\epsilon_{2} +\epsilon_{4} & g_{3,4}  \\
	g_{1,4} & g_{2,4} & g_{3,4} & \epsilon_{1} +\epsilon_{2} +\epsilon_{3} 
\end{bmatrix}
\end{align}
and
\begin{align}
	\operatorname{H}_{\scriptscriptstyle{2}}^{\scriptscriptstyle{(4)}}=\begin{bmatrix}
		\epsilon _3+\epsilon _4 & g_{2,3} & g_{1,3} & g_{2,4} & g_{1,4} & 0  \\
		g_{2,3} & \epsilon _2+\epsilon _4 & g_{1,2} & g_{3,4} & 0 & g_{1,4} \\
		g_{1,3} & g_{1,2} & \epsilon _1+\epsilon _4 & 0 & g_{3,4} & g_{2,4} \\
		g_{2,4} & g_{3,4} & 0 & \epsilon _2+\epsilon _3 & g_{1,2} & g_{1,3}  \\
		g_{1,4} & 0 & g_{3,4} & g_{1,2} & \epsilon_1+\epsilon _3 & g_{2,3} \\
		0 & g_{1,4} & g_{2,4} & g_{1,3} & g_{2,3} & \epsilon _1+\epsilon _2 
	\end{bmatrix}	
\end{align}
In addition, within the dynamics generated by (\ref{Fock-exe2:H}) it is not restrictive to treat the $k=1$ qubit as {\textquotedblleft}central{\textquotedblright}, i.e. as open subsystem, and to investigate its survival probability in contact with the environment defined by the other $N-1$ qubits, with $N$ arbitrary large. 

\section{Survival probability}
\label{sec:sp}

We surmise that the parameters in (\ref{Fock-exe2:H}) yield a non-degenerate spectrum $\left\{ \epsilon_{i} \right\}_{i=1}^{M}$ in one to one correspondence with eigenstates $\left\{ \phi_{i} \right\}_{i=1}^{M}$  with 
\begin{align}
	M=2^{N}
\end{align}
The evolution of an arbitrary initial state $\psi$ then reads
\begin{align}
	\operatorname{U}(t)\psi=\sum_{i=1}^{M}\phi_{i}\,e^{-\imath
\,\epsilon_{i}\,t}\left \langle\,\phi_{i}\,,\psi\,\right\rangle
\end{align}
The survival probability amplitude of the state $\psi$ is 
\begin{align}
	\left \langle\,\psi\,,	\operatorname{U}(t)\psi\,\right\rangle=\sum_{i=1}^{M}\,e^{-\imath\,\epsilon_{i}\,t}\left |\left \langle\,\phi_{i}\,,\psi\,\right\rangle\right |^{2}
\end{align}
If we denote by  
\begin{align}
	c_{i}^{\scriptscriptstyle{(\psi)}}=\left |\left \langle\,\phi_{i}\,,\psi\,\right\rangle\right |^{2}
\end{align}
the probability that an energy measurement of the initial state yields $\epsilon_{i}$ as an outcome,  the survival probability of the state $\psi$ at time $t$ is
\begin{align}
	&	\wp(t):=\left |\left \langle\,\psi\,,\operatorname{U}(t)\psi\,\right\rangle\right |^{2}=\sum_{i,j=1}^{M} \cos(\epsilon_{i,j}\,t)
	\,c_{i}^{\scriptscriptstyle{(\psi)}}\,c_{j}^{\scriptscriptstyle{(\psi)}}
	&& \epsilon_{i,j}:=\epsilon_{i}-\epsilon_{j}
	\label{sp:sp}
\end{align}
Elementary trigonometry and probability conservation yield the equivalent expression
\begin{align}
	\wp(t)=1-4\sum_{i=1}^{M}\sum_{j=i+1}^{M} \sin^{2}\frac{\epsilon_{i,j}\,t}{2}\,c_{i}^{\scriptscriptstyle{(\psi)}}\,c_{j}^{\scriptscriptstyle{(\psi)}}
\end{align}

\subsection{The survival probability of an excitation of a multi-qubit system is an almost periodic function}
\label{sec:sp-ap}

It is expedient to recall that
\begin{definition}{(\textbf{Harald Bohr})}
	A continuous function $ f\colon\mathbb{R}\mapsto\mathbb{R} $  is called \emph{almost periodic} if, for every $ \varepsilon\,>\,0 $, there exists a $ T_{\varepsilon} $ such that for every $ t $ the  interval $ [t\,,t+T_{\varepsilon}) $  contains at least one $ \tau $ such that
	\begin{align}
		|f(t)-f(t+\tau)|\,<\,\varepsilon
	\end{align}
\end{definition}
\noindent In other words, an almost periodic function after a sufficiently long time will return arbitrarily close to any value it has taken. Cardinal results of the theory of almost periodic functions \cite{CorC2009} are that any trigonometric polynomial
\begin{align}
	f(t)=\sum_{n=0}^{N} A_{n}\,e^{\imath\,\lambda_{n}\,t} 
\end{align}
with $A_{n}$'s complex valued and $\lambda_{n}$  real valued is almost periodic, and the same holds for any  absolutely converging Fourier series:
\begin{align}
&	f(t)=\sum_{n=0}^{\infty} A_{n}\,e^{\imath\,\lambda_{n}\,t} && \& &\sum_{n=0}^{\infty} |A_{n}|\,<\,\infty
\end{align}
Here, it is worth recalling that absolute convergence also implies convergence in the squared Euclidean norm, i.e., in the Besicovitch sense, \cite{CorC2009}. The physical expected consequence is that when $N$ tends to infinity, the absolute convergence of the survival amplitude implies that of the survival probability. 

We can immediately conclude that the survival probability of any quantum state evolving according to a unitary transformation generated by a Hamilton operator with a pure point spectrum is an almost periodic function. The recurrence time can, however, be arbitrarily long for systems with large or countable states. These are the questions that experiment designed in \cite{KaWuClPe2026} set out to investigate for a multi-qubit platform. 

\section{Survival probability of a local excitation of the solvable qubit chain}
\label{sec:sp-explicit}

\begin{figure}[H]
	\begin{subfigure}{.5\textwidth}
		\centering
		\includegraphics[width=\textwidth]{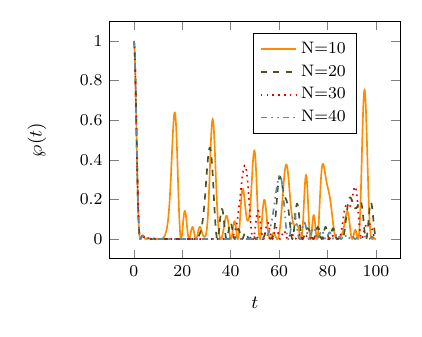}
		\caption{\label{fig:explicit1}}
	\end{subfigure}%
	\begin{subfigure}{.5\textwidth}
		\centering
		\includegraphics[width=\textwidth]{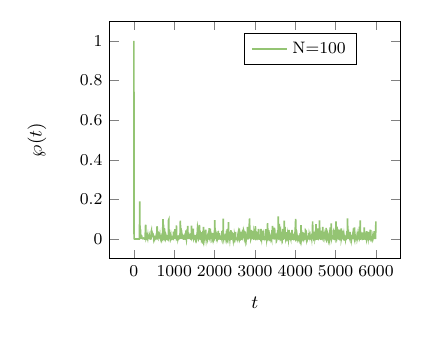}
		\caption{\label{fig:explicit2}}
	\end{subfigure}
	\caption{(\ref{fig:explicit1}): Evolution of the survival probability (\ref{sp-explicit:sp}) (ordinate) versus time (abscissa) for $N=10,20,30,40$ qubits.  For short times all curves are superimposed. (\ref{fig:explicit2}): in the solvable chain with $N=100$ no revival occur over the plotted time scale.  All plots for $g=1/\sqrt{2}$. \label{fig:explicit}}
\end{figure}

Let us analyze the fate of the survival probability of single qubit excitation for the chain described in section~\ref{sec:Fock-exe1}.  
Specifically, we set
\begin{align}
	\psi=\operatorname{a}_{1}^{\dagger}\Phi_{0}
\end{align}
This state has a non-vanishing projection only on the one Bogolyubov-Valatin particle sector. The survival amplitude is therefore a trigonometric polynomial with only $N$ terms:  
\begin{align}
	\left \langle\,\psi\,,\operatorname{U}(t)	\psi\,\right\rangle=e^{-\imath\,\omega\,t}\sum_{\ell=1}^{N} e^{-\imath\,2\,g\,t\,\cos\frac{\ell\, \pi}{N+1}}\frac{\sin^{2} \frac{\ell\,\pi}{N+1}}{Z}
\end{align}
It is important to note that a local excitation has a non-vanishing projection on all the eigenstates of the Bogolyubov--de Gennes matrix. This is the basic mechanism responsible for the equilibration and thermalization in large systems \cite{BaGrVuZa2025,CaBaLuMGVu2025}. The corresponding expression of the survival probability is
\begin{align}
	\wp(t)=\sum_{\ell,k=1}^{N} \cos\left(2\,g\,t\,\left (\cos\frac{\ell\, \pi}{N+1}-\cos\frac{k\, \pi}{N+1}\right )\right)\frac{\sin^{2} \frac{\ell\,\pi}{N+1}}{Z}\frac{\sin^{2} \frac{k\,\pi}{N+1}}{Z}
	\label{sp-explicit:sp}
\end{align} 
or equivalently
\begin{align}
	\wp(t)=1-4\sum_{\ell=1}^{N} \sum_{k>\ell}^{N} \sin^{2}\left(g\,t\,\left (\cos\frac{\ell\, \pi}{N+1}-\cos\frac{k\, \pi}{N+1}\right )\right)\frac{\sin^{2} \frac{\ell\,\pi}{N+1}}{Z}\frac{\sin^{2} \frac{k\,\pi}{N+1}}{Z}
\end{align}
From the second expression, we see that for timescales 
\begin{align}
	t \,\sqrt{\operatorname{Var}_{\operatorname{P}_{\psi}}\left (\operatorname{H}_{\scriptscriptstyle{1}}^{\scriptscriptstyle{(N)}}\right )}\,\lesssim\,O(1)
\end{align}
where
\begin{align}
	\operatorname{Var}_{\operatorname{P}_{\psi}}\left (\operatorname{H}_{\scriptscriptstyle{1}}^{\scriptscriptstyle{(N)}}\right ):=\left \langle\,\psi\,,\operatorname{H}_{\scriptscriptstyle{1}}^{\scriptscriptstyle{(N)}}{}^{2}\,\psi\,\right\rangle-\big{(}\left \langle\,\psi\,,\operatorname{H}_{\scriptscriptstyle{1}}^{\scriptscriptstyle{(N)}}\,\psi\,\,\right\rangle\big{)}^{2}
\end{align} 
the deviations of the survival probability from its initial value equal to one  are of the order of the square of the intensity of the qubit coupling constant $g$ for small values of it . Indeed, a straightforward calculation gives
\begin{align}
	\operatorname{Var}_{\operatorname{P}_{\psi}}\left (\operatorname{H}_{\scriptscriptstyle{1}}^{\scriptscriptstyle{(N)}}\right )=4\,g^{2}\sum_{\ell=1}^{N}\cos^{2} \left(\frac{\ell\,\pi}{N+1}\right)\frac{\sin^{2} \frac{\ell\,\pi}{N+1}}{Z}=g^{2}
\end{align}
In Fig~\ref{fig:explicit}  we show the typical behavior of the survival probability for a value of the qubit for chains of different numbers $N$ of qubits. 

\subsection{Universal relaxation behavior before revival}

\begin{figure}
		\centering
		\includegraphics[width=0.7\textwidth]{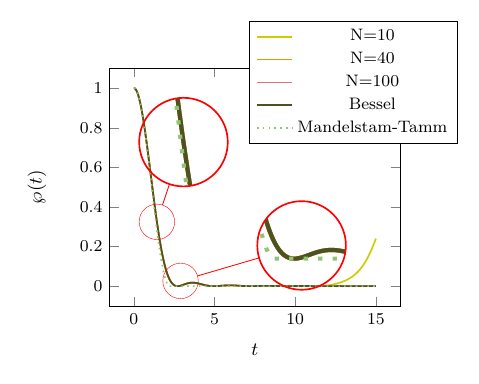}
	\caption{Evolution of the survival probability (\ref{sp-explicit:sp}) (ordinate) versus time (abscissa) for $N=10,20,30,40,100$ contrasted with the asymptotic expression specified by the Bessel function (\ref{sp-explicit:Bessel}). For short times all curves are superimposed.  The dotted curve is the prediction of the Mandelstam-Tamm lower bound (\ref{Zeno:lb}) set identically to zero for times larger than the threshold values (\ref{Zeno:splim}).  For times large $g t\,\geq\,10$  the curve for $N=10$ differs from the others, in that it displays a first revival.  All plots for $g=1/\sqrt{2}$}
	\label{fig:chain_universal}
\end{figure}

A remarkable feature of Fig.~\ref{fig:explicit1} is that the survival probability curves for chains of length $N=10$ and larger appear to overlap before the first revival. The time at which the first revival occurs is strongly dependent on $N$ (see Section~\ref{sec:Kac}).
To shed light on the apparent pre-revival universal behavior, we consider the limit of infinite chain and approximate sums with integrals. 
In this {\textquotedblleft}continuum{\textquotedblright} limit, the survival probability becomes 
\begin{align}
	\wp(t) \simeq\frac{4}{\pi^{2}}\left |\int_{0}^{\pi}\mathrm{d}x e^{2\,\imath\,g\,t\,\cos x}
	\sin^{2}x\right |^{2}=\left |\frac{\mathcal{J}_{\scriptscriptstyle{1}}(2\,g\,t)}{g\,t}\right |^{2}
	\label{sp-explicit:Bessel}
\end{align}
where  $\mathcal{J}_{\scriptscriptstyle{1}}$ is the Bessel function of the first kind. Hence, the continuum limit approximation
decays exponentially fast for large times
\begin{align}
	\wp(t) \overset{t\uparrow \infty}{\sim}\frac{1 }{\pi \,(t \,g)^{3}}
	\cos^{2} \left(2\, g \,t+\frac{\pi }{4}\right)
\end{align}
Fig.~\ref{fig:chain_universal} shows how the continuum limit captures the universal behavior of the survival probability before the first revival. 
In summary, the model constitutes an example in the context of unitary dynamics of how an irreversible evolution may very accurately reproduce the behavior of an almost periodic dynamics before a revival.

\section{Survival probability of a local excitation of the general multi-qubit model}
\label{sec:mqm}

\begin{figure}
	\begin{subfigure}{.5\textwidth}
		\centering
		\includegraphics[width=1.05\textwidth]{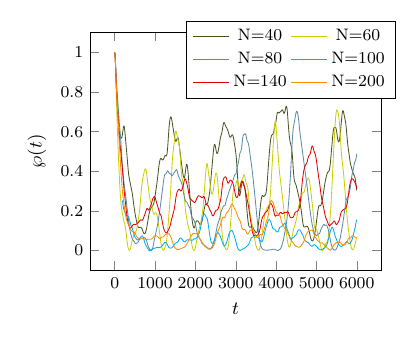}
		\caption{\label{fig:Hexp1}}
	\end{subfigure}%
		\begin{subfigure}{.5\textwidth}
		\centering
		\includegraphics[width=\textwidth]{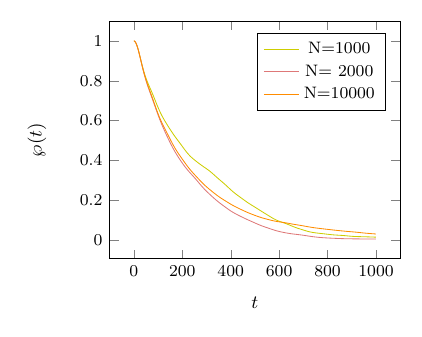}
		\caption{\label{fig:Hexp3}}
	\end{subfigure}%
	\caption{(\ref{fig:Hexp1}) \& (\ref{fig:Hexp3}): time evolution of the  survival probability (ordinate) of a single particle (central qubit) excitation for the dynamics generated by \eqref{Fock-exe2:H} versus time (abscissa). As the number of particles $N$ increases oscillations are suppressed and the survival probability appears to decay exponentially (see section~\ref{sec:mqm-Lee}). The level splittings of all qubits but the central one are sampled according to identical independent uniform distributions in $[\omega-\Delta,\omega+\Delta]$.   The mean value $\omega$ is exactly the splitting of the central qubit. In numerics we measure all energies in units of $\omega$, and we set $\omega=1.0$  and $\Delta=0.1$ as in \cite{KaWuClPe2026}. Off diagonal elements	are Gaussian independent identically distributed random variables with mean zero and variance $\sigma^{2}$. In the numerics, we choose $\sigma=\sqrt{1.5\,\Gamma\, \Delta}$ with $\Gamma=10^{-3}$. The plots describe one realization of the random dynamics. We choose the numerical value of $\sigma$ such to obtain curves reproducing the behavior of those of \cite{KaWuClPe2026} where off- diagonal elements of the single particle Hamiltonian are also sampled according to a uniform distribution.	\label{fig:Hexp}	}
\end{figure}

In general, the spectral decomposition of the Hamilton operator (\ref{Fock-exe2:H}) can only be obtained numerically.
In Fig.~\ref{fig:Hexp} we show some typical behavior of the survival probability for the values of the couplings and the qubit level splittings that correspond to the experimental design proposed in \cite{KaWuClPe2026}. In particular, the non dimensional coupling between qubits are of the order $O(10^{-3})$. As a consequence, the single particle Hamiltonian is always positive definite in the parametric range of interest.

In order to delve into the meaning of the numerical results, it  is expedient to summarize some general analytic properties of the survival probability for a many qubit system.

Conservation of the number of quanta (\ref{Fock:quanta}) allows us to compute the survival probability of a local excitation by considering the single particle sector alone. Similarly, the survival probability of an initial state with exactly $n$-quanta is governed by the restriction of the many-body Hamilton operator to the $n$-particle sector. We focus here on the $n=1$ case. The restricted dynamics is that of a $N$-level system:
 \begin{equation}
 	\label{mqm:ode}
 	\begin{split}
 		&			\imath\,\partial_{t}\operatorname{K}(t)=\operatorname{H}_{\scriptscriptstyle{1}}^{\scriptscriptstyle{(N)}}\operatorname{K}(t)
 		\\
 		&			\operatorname{K}_{0}=\operatorname{1}_{N}
 	\end{split}
 \end{equation}
We surmise that the local excitation corresponds to initially preparing the system in the state
\begin{align}
	\psi=e_{1}
	\label{mqm:local}
\end{align}
where $e_{1}$ is the first element of the computational basis of $\mathbb{C}^{N}$.  This choice suggests the block representation 
 \begin{align}
 	\operatorname{H}_{\scriptscriptstyle{1}}^{\scriptscriptstyle{(N)}}=
 	\begin{bmatrix}
 		\omega	& g^{\dagger} \\ g & \Omega  
 	\end{bmatrix}
 	\label{mqm:H}
 \end{align}
 where
 \begin{itemize}
 	\item $\omega$ is the level splitting of the qubit initially in the excited state when the remaining $N-1$ are in the ground state.
 	\item $g\,\in\,\mathbb{C}^{N}$ is a vector that encodes the interaction between the initially excited qubit and the surrounding qubits.
 	\item $\Omega$ is a self-adjoint, $N-1\,\times\,N-1$ square matrix whose diagonal elements specify the level splittings of the surrounding qubits. The off-diagonal entries describe the intensity of the couplings between the environment qubits.
 	As a consequence, when modeling $\operatorname{H}_{\scriptscriptstyle{1}}^{\scriptscriptstyle{(N)}}$ with a random matrix ensemble, the statistics of diagonal entries should differ from that of off-diagonal ones to be physically descriptive. The former describe qubit energy splittings while the latter model interactions. 
 \end{itemize}
As customary, we solve (\ref{mqm:ode}) by introducing the \emph{resolvent} operator
 \begin{align}
 &	\operatorname{R}(z)=-\imath\int_{0}^{\infty}\mathrm{d}t\,e^{\imath\,z\,t}\,\operatorname{K}(t) && \forall\,\operatorname{Im}z,t>0
 	\label{mqm:Rdef}
 \end{align} 
 Mathematically, the resolvent turns the  system of differential equations into an algebraic problem whose solution admits the straightforward expression
 \begin{align}
 	\operatorname{R}(z)=(z\,\operatorname{1}_{N}-\operatorname{H}_{\scriptscriptstyle{1}}^{\scriptscriptstyle{(N)}})^{-1}
 \end{align}
To analyze the evolution of the central qubit, we need to decompose the resolvent into blocks corresponding to the partition (\ref{mqm:H}) of the one-particle sector Hamiltonian.

 Upon combining the well known matrix block inversion and Sherman-Morrison formulae (see appendix~\ref{app:LA}), we obtain
 \begin{align}
 	\operatorname{R}(z)=
 	\begin{bmatrix}
 		\dfrac{1}{z-\omega-\Lambda(z)} 
 		& \phantom{g^{\dagger}\operatorname{R}_{\Omega}(z}&
 		\dfrac{g^{\dagger}\operatorname{R}_{\Omega}(z)}{z-\omega-\Lambda(z)}
 		\\[0.3cm]
 	   \dfrac{\operatorname{R}_{\Omega}(z)\,g}{z-\omega-\Lambda(z)} 
 	    &  \phantom{g^{\dagger}\operatorname{R}_{\Omega}(z} & 
 	  \operatorname{R}_{\Omega}(z)
 	   +\dfrac{\operatorname{R}_{\Omega}(z)\,g\,g^{\dagger}\operatorname{R}_{\Omega}(z)}{z-\omega-\Lambda(z)}
 	\end{bmatrix}
 	\label{mqm:res}
 	\end{align}
 	where we introduce the {\textquotedblleft}level-shift{\textquotedblright} \cite{CoTaDuRoGr1998} scalar function
 	\begin{align}
 		\Lambda(z)=\left \langle\,g\,,(z\,\operatorname{1}_{N-1}-\Omega )^{-1}g\,\right\rangle
 		\label{mqm:ls}
 	\end{align}
 	and the $N-1\,\times\,N-1$ matrix describing the reduced resolvent of the environment dynamics:
 	\begin{align}
 	\operatorname{R}_{\Omega}(z)= 	(z\,\operatorname{1}_{N-1}-\Omega)^{-1}
 		\label{mqm:Renv}
 	\end{align}
 	The survival probability of the central qubit, coincides with the Laplace anti-transform of the upper diagonal element of the resolvent
 	\begin{align}
 	\wp(t)=	\left |\,\,\int\limits_{\mathbb{R}+\imath\,\varepsilon}\frac{\mathrm{d}z}{2\,\pi}\frac{e^{-\imath\,z\,t}}{z-\omega -\Lambda(z)}\right |^{2}
 	\label{mqm:sp}
 	\end{align}

 \subsection{Perturbative expression of the survival probability}
 \label{sec:mqm-pt}
 
For a system with a finite number $N$ of qubits, (multiscale) perturbation theory offers some insight into the evolution of the survival probability.
 
\begin{figure}[H]
	\begin{subfigure}{.5\textwidth}
		\centering
		\includegraphics[width=1.1\textwidth]{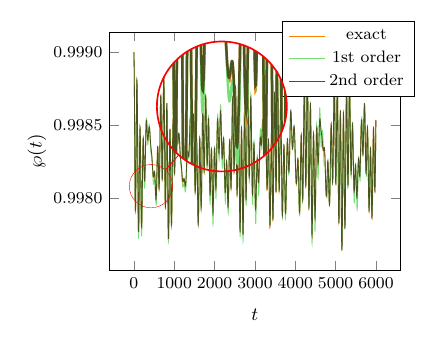}
		\caption{Survival probability for $N=8$\label{fig:pt1}}
	\end{subfigure}
	\begin{subfigure}{.5\textwidth}
		\centering
		\includegraphics[width=\textwidth]{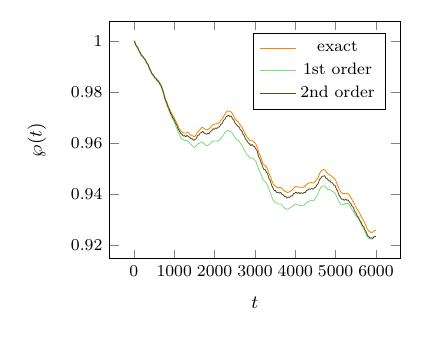}
		\caption{Survival probability for $N=80$ \label{fig:pt2}}
	\end{subfigure}
	\caption{Evolution of the survival probability (ordinate) versus time (abscissa). In both (\ref{fig:pt1}) \& (\ref{fig:pt2}) $\sigma=\sqrt{1.5\,\Gamma\, \Delta}/10$ whereas all other parameters are as in Fig~\ref{fig:Hexp}. 	As expected, increasing the number of qubits at fixed value of the coupling constant  decreases the accuracy of standard perturbation theory \label{fig:pt}}
	\end{figure}

In this regime, it is expedient  to decompose the single particle Hamilton operator into diagonal and interaction part
  \begin{align}
  	\operatorname{H}_{\scriptscriptstyle{1}}^{\scriptscriptstyle{(N)}}=\operatorname{H}_{\scriptscriptstyle{1}}^{\scriptscriptstyle{(N,0)}}+\varepsilon\,\operatorname{V}^{\scriptscriptstyle{(N)}}
  \end{align}
  where $ \operatorname{H}_{\scriptscriptstyle{1}}^{\scriptscriptstyle{(N,0)}}$ and $\operatorname{V}^{\scriptscriptstyle{(N)}}$ are a diagonal, and a {\textquotedblleft}hollow{\textquotedblright} (i.e. with zeroes on the main diagonal) $N\,\times\,N$ self-adjoint matrices, respectively. 
  The non-dimensional parameter $\varepsilon$ controls the strength of the interaction. If we now introduce the projector on subspace of the  one-particle Hilbert space transversal to the span of the initial state (\ref{mqm:local})
  \begin{align}
  	\operatorname{Q}=\sum_{i=2}^{N}e_{i}e_{i}^{\dagger}
  \end{align}
  we readily obtain
  \begin{align}
  	\begin{split}
  		&	0\oplus g=\varepsilon\,\operatorname{Q}\operatorname{V}^{\scriptscriptstyle{(N)}}e_{1}
  		\\
  		&0\oplus \Omega=\operatorname{Q}\operatorname{H}_{\scriptscriptstyle{1}}^{\scriptscriptstyle{(N)}}\operatorname{Q}
 	\end{split}
  \end{align}
  These identities allow us to recast the level shift function into the form
  \begin{align}
  	\Lambda(z)=\varepsilon^{2}\left \langle\,e_{1}\,,\operatorname{V}^{\scriptscriptstyle{(N)}}\operatorname{Q}(z\,\operatorname{1}_{N}-\operatorname{H}_{\scriptscriptstyle{1}}^{\scriptscriptstyle{(N,0)}}-\varepsilon\,\operatorname{Q}\operatorname{V}^{\scriptscriptstyle{(N)}}\operatorname{Q})^{-1}\operatorname{Q}\operatorname{V}^{\scriptscriptstyle{(N)}}e_{1}\,\right\rangle
  \end{align}  
 By construction, $\operatorname{Q}$ is idempotent and commutes with $ \operatorname{H}_{\scriptscriptstyle{1}}^{\scriptscriptstyle{(N,0)}}$.
We avail us of these properties to expand the level shift in powers of $\varepsilon$. A straightforward expansion 
of the level shift operator yields 
  \begin{align}
  &	(z\,\operatorname{1}_{N}-\operatorname{H}_{\scriptscriptstyle{1}}^{\scriptscriptstyle{(N,0)}}-\varepsilon\,\operatorname{Q}\operatorname{V}^{\scriptscriptstyle{(N)}}\operatorname{Q})^{-1}
  	=\left(z\,\operatorname{1}_{N}-\operatorname{H}_{\scriptscriptstyle{1}}^{\scriptscriptstyle{(N,0)}}\right)^{-1}
  \nonumber\\
  &+\varepsilon\,\left(z\,\operatorname{1}_{N}-\operatorname{H}_{\scriptscriptstyle{1}}^{\scriptscriptstyle{(N,0)}}\right)^{-1}\operatorname{Q}\operatorname{V}^{\scriptscriptstyle{(N)}}\operatorname{Q}\left(z\,\operatorname{1}_{N}-\operatorname{H}_{\scriptscriptstyle{1}}^{\scriptscriptstyle{(N,0)}}\right)^{-1}
  \nonumber\\
&  +\varepsilon^{2}\,\left(z\,\operatorname{1}_{N}-\operatorname{H}_{\scriptscriptstyle{1}}^{\scriptscriptstyle{(N,0)}}\right)^{-1}\operatorname{Q}\operatorname{V}^{\scriptscriptstyle{(N)}}\operatorname{Q}\left(z\,\operatorname{1}_{N}-\operatorname{H}_{\scriptscriptstyle{1}}^{\scriptscriptstyle{(N,0)}}\right)^{-1}\operatorname{Q}\operatorname{V}^{\scriptscriptstyle{(N)}}\operatorname{Q}\left(z\,\operatorname{1}_{N}-\operatorname{H}_{\scriptscriptstyle{1}}^{\scriptscriptstyle{(N,0)}}\right)^{-1}+\dots
  \end{align}
  In the absence of symmetries, a generic $\operatorname{H}_{\scriptscriptstyle{1}}^{\scriptscriptstyle{(N)}}$ has a non-degenerate spectrum. Nevertheless, the above expansion produces poles of order higher than the first to spurious secular terms in the expansion of the probability amplitude:
  \begin{align}
  	\alpha(t)=\int\limits_{\mathbb{R}+\imath\,\varepsilon}\frac{\mathrm{d}z}{2\,\pi}\frac{e^{-\imath\,z\,t}}{z-\omega -\Lambda(z)}
  	\label{mqm-pt:pa}
  \end{align}
  This is a well-known pathology of Taylor expansions of bounded functions that can be cured by multiscale perturbation theory; see, e.g. \cite{GenG2010}. In the case of (\ref{mqm-pt:pa}) the procedure reduces to an expansion  with respect to a {\textquotedblleft}renormalized{\textquotedblright} diagonal Hamiltonian $\operatorname{H}_{\scriptscriptstyle{1}}^{\scriptscriptstyle{(N,R)}}$ satisfying
  \begin{align}
	\operatorname{H}_{\scriptscriptstyle{1}}^{\scriptscriptstyle{(N,0)}}=\operatorname{H}_{\scriptscriptstyle{1}}^{\scriptscriptstyle{(N,R)}}+\delta\operatorname{H}_{\scriptscriptstyle{1}}^{\scriptscriptstyle{(N,R)}}
  	\label{counter}
  \end{align}
  We then treat {\textquotedblleft}counter-term{\textquotedblright} $\delta\operatorname{H}_{\scriptscriptstyle{1}}^{\scriptscriptstyle{(N,R)}}$ as a perturbation whose value we determine order by order by subtracting residues of non-single poles generated by the expansion.  Finally, we use  (\ref{counter}) to reconstruct $\operatorname{H}_{\scriptscriptstyle{1}}^{\scriptscriptstyle{(N,R)}}$ with the desired accuracy.
 Due to cancelations, the leading order expression of the expansion does not require multi-scale analysis and yields the simple expression, already given in \cite{KaWuClPe2026}
  \begin{align}
  	&	\wp(t)=1
  	-4 \,\varepsilon^{2}\sum_{j\neq 1}\sin^{2}\left(\frac{\epsilon_{1,j}^{\scriptscriptstyle{(0)}}t}{2}\right)\,\frac{|\operatorname{V}_{1,j}^{\scriptscriptstyle{(N)}}|^{2}}{\epsilon_{1,j}^{\scriptscriptstyle{(0)}}{}^{2}}+O(\varepsilon^{3})
  	\label{mqm-pt:sp}
  \end{align} 
  where we define
  \begin{align}
  &	\epsilon_{i}^{\scriptscriptstyle{(0)}}=\left \langle\,e_{i}\,,\operatorname{H}_{\scriptscriptstyle{1}}^{\scriptscriptstyle{(N,0)}}e_{i}\,\right\rangle
  && i=1,\dots,N
  \end{align}
  with in particular $\epsilon_{1}^{\scriptscriptstyle{(0)}}=\omega $ and
  \begin{align}
  	\begin{split}
  		&  \epsilon_{i,j}^{\scriptscriptstyle{(0)}}:=\epsilon_{i}^{\scriptscriptstyle{(0)}}-\epsilon_{j}^{\scriptscriptstyle{(0)}}
  		\\
  		&\operatorname{V}_{1,j}^{\scriptscriptstyle{(N)}}:=\left \langle\,e_{i}\,,\operatorname{V}^{\scriptscriptstyle{(N)}}e_{j}\,\right\rangle
  	\end{split}
  	\label{mqm-pt:sp-pt}
  \end{align}
  The renormalization procedure becomes necessary to remove secular terms occurring at  $O(\varepsilon^{4})$. The
corresponding expression of the survival probability is cumbersome, and we write it in Appendix~\ref{app:hopt}.
  We compare both (\ref{mqm-pt:sp}), and the $O(\varepsilon^{4})$ refined approximation (\ref{hopt:sp}) with the non-perturbative numerical predictions  of Fig~\ref{fig:pt}. The interest in the comparison lies in providing a reference for the interpretation of experimental data as the platforms \cite{WuYaAnAnCle2024,AnAbMi2025} allow one to tune the strength of qubit interactions.

 \subsection{Survival probability for an infinite environment}
 \label{sec:mqm-Lee}

Now, let us suppose that the self-adjoint matrix $\Omega $ in (\ref{mqm:H}) is diagonal. In principle, this is not  a restrictive assumption, as we can imagine performing a partial diagonalization at the price of redefining the coupling vector $g$. 

Under this hypothesis, the level shift function (\ref{mqm:ls}) reduces to
\begin{align}
	\Lambda(z)=\sum_{\ell=1}^{N-1}\frac{|g_{\ell}|^{2}}{z-\Omega_{\ell,\ell}}
	\label{Lee}
\end{align}
We now formulate the following additional hypotheses on the entries of the single particle Hamiltonian.
\begin{enumerate}[label={\upshape\bfseries H\arabic*}]
	\item \label{H1} We fix the energy splitting  $\omega$ of the central qubit  to a constant deterministic value.
	\item \label{H2} We assume that the diagonal matrix $\Omega$ has non-vanishing entries of the form
	\begin{align}
&		\Omega_{\ell,\ell}=\omega+\xi_{\ell} && \ell=1,\dots,N-1
	\end{align}
	where the $\xi_{\ell}$'s are independent identically and uniformly distributed random variable taking values in $[-\Delta\,,\Delta]$:
	\begin{align}
		\operatorname{Prob}(\xi_{\ell}\in [a,b] \subset [-\Delta,\Delta])=\frac{b-a}{2\,\Delta}
	\end{align}
	\item\label{H3} Finally, we suppose that in (\ref{mqm:H}) the coupling vector $g$ is a $N-1$ has real entries specified by independent identically distributed Gaussian random variables with mean zero and variance: 
	\begin{align}
		\sigma_{N}^{2}=\frac{\sigma^{2}}{N}
	\label{mqm-Lee:variance}
	\end{align}
\end{enumerate} 
We are interested in the mean value of the survival probability
   \begin{align}
&   	\operatorname{E}\wp_{N}(t)=\operatorname{E}\left |\,\,\int\limits_{\mathbb{R}+\imath\,\varepsilon}\frac{\mathrm{d}z}{2\,\pi}\frac{e^{-\imath\,z\,t}}{z-\omega -\sum_{\ell=1}^{N-1}\frac{g_{\ell}^{2}}{z-\omega-\xi_{\ell}}}\right |^{2} && \varepsilon>0
   \label{mqm-Lee:sp}
   \end{align}
   
\subsubsection{Evaluation of the mean value of the survival probability}
\label{sec:mqm-Lee-mean}

By (\ref{mqm:Rdef}),  we can certainly use Schwinger's representation
\begin{align}
&	\frac{\imath}{A}=\int_{0}^{\infty}\mathrm{d}s\,e^{\imath\,A\,s} && \mbox{for}&\operatorname{Im}A>0
\end{align}
 to evaluate the Gaussian integrals over the random couplings $g_{i}$'s in (\ref{mqm-Lee:sp}) according to~\ref{H3}. We get
	\begin{align}
	&\operatorname{E}\wp_{N}(t)=
	\nonumber\\
&	\int_{0}^{\infty}\mathrm{d}s_{\mathfrak{1}}\int_{0}^{\infty}\mathrm{d}s_{\mathfrak{2}}\int\limits_{\mathbb{R}+\imath\varepsilon}\frac{\mathrm{d}z_{\mathfrak{1}}}{2\,\pi}\int\limits_{\mathbb{R}-\imath\varepsilon}\frac{\mathrm{d}z_{\mathfrak{2}}}{2\,\pi}e^{\imath\,\sum\limits_{\ell=\mathfrak{1}}^{\mathfrak{2}}(-)^{\ell}\big{(}z_{\ell}\,t-s_{\ell}\, (z_{\ell}-\omega)\big{)}}F_{N}(z_{\mathfrak{1}},z_{\mathfrak{2}},s_{\mathfrak{1}},s_{\mathfrak{2}})
\label{mqm-Lee:Ewp}
\end{align}
with
\begin{align}
	F_{N}(z_{\mathfrak{1}},z_{\mathfrak{2}},s_{\mathfrak{1}},s_{\mathfrak{2}})=\left(\operatorname{E}\frac{1}{\sqrt{1+2\,\sigma_{N}^{2}\,\imath\left(\frac{s_{\mathfrak{1}}}{z_{\mathfrak{1}}-\omega-\xi}-\frac{s_{\mathfrak{2}}}{z_{\mathfrak{2}}-\omega-\xi}\right)}}\right)^{N-1}
\end{align}
Next, we use (\ref{mqm-Lee:variance}) and take the limit of infinite environment. We obtain
\begin{align}
	\lim_{N\uparrow \infty}F_{N}(z_{\mathfrak{1}},z_{\mathfrak{2}},s_{\mathfrak{1}},s_{\mathfrak{2}})=\exp\left(\sigma^{2}\,\imath\operatorname{E}\left(\frac{s_{\mathfrak{1}}}{\xi+\omega-z_{\mathfrak{1}}}-\frac{s_{\mathfrak{2}}}{\xi+\omega -z_{\mathfrak{2}}}\right)\right)
\end{align}
The upshot is
\begin{align}
	\wp(t):=\lim_{N\nearrow\infty}\operatorname{E}\wp_{N}(t)=\left |\,\,\int\limits_{\mathbb{R}+\imath\,\varepsilon}\frac{\mathrm{d}z}{2\,\pi}\frac{e^{-\imath\,z\,t}}{z-\omega +\sigma^{2} \operatorname{E}\frac{1}{\xi+\omega-z}}\right |^{2} && \forall\varepsilon>0
	\label{mqm-Lee:sp-Lee}
\end{align}
where now the expectation value $\operatorname{E}$ is restricted to the prototype $\xi$ of the random variables defined in hypothesis~\ref{H2}. From a mathematical point of view, we recognize that the expectation value recovers the Stieltjes transform \cite[\S~2.3.2]{PoBo2020} of the distribution of the random variable $\xi$.

\subsubsection{Analysis of the limit value of the survival probability} 
There are several important consequences. 
\begin{description}[leftmargin=0pt]
\item[The survival probability does not fluctuate in the limit of infinite environment.] Namely, using \ref{H2} \ref{H3}, the expectation value of the level shift function yields
\begin{align}
	\operatorname{E}\sum_{\ell=1}^{N-1}\frac{g_{\ell}^{2}}{\omega+\xi_{\ell}-z}=\sigma^{2}\operatorname{E}\frac{1}{\xi+\omega-z}
\end{align}
Hence, if we define the fluctuating survival probability amplitude
\begin{align}
	\begin{split}
&	\alpha_{N}(t)= \int\limits_{\mathbb{R}+\imath\,\varepsilon}\frac{\mathrm{d}z}{2\,\pi}\frac{e^{-\imath\,z\,t}}{z-\omega -\sum_{\ell=1}^{N-1}\frac{g_{\ell}^{2}}{z-\omega-\xi_{\ell}}}
	\end{split}
	\label{mqm-Lee:pa-Lee}
\end{align}
we readily verify the chain of identities
\begin{align}
	\wp(t)=\lim_{N\nearrow\infty}\operatorname{E}|\alpha_{N}(t)|^{2}=
	\left |\lim_{N\nearrow\infty}\operatorname{E}\alpha_{N}(t)\right |^{2}:=|\alpha(t)|^{2}
\end{align}
which prove that the variance of the probability amplitude vanishes in the limit. 
 \item[Recovery of the non relativistic Lee model.] If we take into account that
 \begin{align}
 	\operatorname{E}\frac{1}{\xi+\omega-z}=\frac{1}{2\,\Delta}\int_{-\Delta}^{\Delta}\mathrm{d}x\frac{1}{x+\omega-z}
 \end{align}
we recognize that  (\ref{mqm-Lee:pa-Lee}) is exactly the probability amplitude of a non-relativistic Lee model 
with non-dimensional coupling constant
\begin{align}
\varkappa^{2}:=\frac{\sigma^{2}}{2\,\omega\,\Delta}
\label{mqm-Lee:coupling}
\end{align}
determined by the ratio between the variance of the Gaussian distributed off-diagonal elements  $g_{i}$'s and the spread of the uniformly diagonal elements of the Hamiltonian of the single particle sector. We emphasize that if we replace the Gaussian distribution of the off-diagonal elements with a uniform one centered at zero and spread $\tilde{\Delta}$ around the mean value, then we would have obtained
\begin{align}
	\varkappa^{2}:=\frac{\tilde{\Delta}^{2}}{3\,\omega\,\Delta}
\end{align}
The relativistic Lee model \cite{LeeT1954,ArMuKaGo1957}  has been extensively studied as an example of a field theory with an unstable state \cite{FoGhRi1978}. The non-relativistic limit  also provides a convenient mathematical model to exemplify the coexistence of unitary evolution and dissipation and the role of Khalfin's theorem \cite{KhaL1957} for systems with a continuous spectrum \cite{CoTaDuRoGr1998,AnaC2018}.  In particular, we refer to \cite{WolT2013} for a comprehensive study of the non relativistic Lee model for any strength of the {\textquotedblleft}coupling{\textquotedblright} $ \varkappa^{2}$. The same results are also more briefly presented in \cite{JaKrPi2006,DoGoMG2020}. In appendix~\ref{app:Lee} we recall the analytic continuation procedure that provides the quantitative basis to understand the qualitative properties of the evolution of the survival probability that we describe below.
\end{description}

\subsubsection{Qualitative analysis of time evolution of the survival probability}
\label{sec:mqm-Lee-t}
\begin{figure}
	\begin{subfigure}{.5\textwidth}
		\centering
		\begin{tikzpicture}	[background rectangle/.style={fill=platinum}, show background rectangle,thin,	set/.style = {circle, minimum size = 3cm, fill=FigureBlue},baseline={([yshift=-2.5ex]current bounding box.center)}]
			\tikzset{branches/.style={%
					postaction={decorate,decoration={%
							markings,
							mark=at position 0.2 with {\arrow[scale=1,>=Latex]{>}},	
							mark=at position 0.4 with {\arrow[scale=1,>=Latex]{>}},	
							mark=at position 0.6 with {\arrow[scale=1,>=Latex]{>}},
							mark=at position 0.8 with {\arrow[scale=1,>=Latex]{>}},
							mark=at position 0.92 with {\arrow[scale=1,>=Latex]{>}}
						}
			}}};
			\tikzset{midarrow/.style={%
					postaction={decorate,decoration={%
							markings,
							mark=at position 0.35 with {\arrow[scale=1,>=Latex]{<}},
							mark=at position 0.7 with {\arrow[scale=1,>=Latex]{<}}
						}
			}}};
			\draw [-{Latex[length=2mm]}, thick] (-2.5,0) --(2.6,0) coordinate (xaxis);
			\draw [-{Latex[length=2mm]},thick] (0.0,-2.1)coordinate(-yaxis) -- (0.0,1.0) coordinate (yaxis);
			\node [right] at (xaxis) {$\scriptstyle{\operatorname{Re}(z)}$};
			\node [right] at (yaxis) {$\scriptstyle{\operatorname{Im}(z)}$};
			\node  [below right] at (-yaxis) {\textcolor{darkred}{$\mathcal{C}$}};
			\draw[draw=darkred,line width=0.8pt,midarrow] (-2.5,0.2) -- (2.5,0.2);  
			\begin{scope}
				\clip (-2.5,0.2) rectangle (2.5,-2.7);   
				\draw[draw=darkred, fill=PastelRed, fill opacity=0.2, line width=0.8pt,dashed,branches] (0,0.2) circle(2.4);
			\end{scope}
			\draw[draw=yellow, thick] (0.0,0)--(1.0,0);
			\node [below left] {$O$};
			\draw[draw=darkred] (-0.3,0.0) circle(0.1);
			\node  at (-0.3,0.0) {\textcolor{darkred}{$\times$}};		
			\draw[draw=darkred] (1.4,0.0) circle(0.1);
			\node  at (1.4,0.0) {\textcolor{darkred}{$\times$}};		
		\end{tikzpicture}
		\begin{minipage}{0.95\textwidth}		 
			\caption{Contour on the first Riemann sheet. The analytic extension of the Laplace transform to the lower half plane permits to close the contour at infinity. The non-vanishing contribution to the line integral on the closed contour comes from the contour section parallel to the real axis. By Cauchy's theorem, the line integral must be also equal to the sum over the residues of the poles on the real axis plus the contribution of the line integral encircling the cut on the first Riemann sheet, see \eqref{sp-Lee:Riemann1} . \label{fig:contour1}}
		\end{minipage}
	\end{subfigure}
	\begin{subfigure}{0.5\textwidth}
		\begin{tikzpicture}	[background rectangle/.style={fill=platinum}, show background rectangle,thin,	set/.style = {circle, minimum size = 3cm, fill=FigureBlue},baseline={([yshift=-2.5ex]current bounding box.center)}]
			\tikzset{branches/.style={%
					postaction={decorate,decoration={%
							markings,
							mark=at position 0.2 with {\arrow[scale=1,>=Latex]{>}},	
							mark=at position 0.4 with {\arrow[scale=1,>=Latex]{>}},	
							mark=at position 0.6 with {\arrow[scale=1,>=Latex]{>}},
							mark=at position 0.8 with {\arrow[scale=1,>=Latex]{>}},
							mark=at position 0.92 with {\arrow[scale=1,>=Latex]{>}}
						}
			}}};
			\tikzset{midarrow/.style={%
					postaction={decorate,decoration={%
							markings,
							mark=at position 0.3 with {\arrow[scale=1,>=Latex]{>}},
							mark=at position 0.6 with {\arrow[scale=1,>=Latex]{>}},
							mark=at position 0.9 with {\arrow[scale=1,>=Latex]{>}}
						}
			}}};
			
			\tikzset{rmidarrow/.style={%
					postaction={decorate,decoration={%
							markings,
							mark=at position 0.35 with {\arrow[scale=1,>=Latex]{<}},
							mark=at position 0.7 with {\arrow[scale=1,>=Latex]{<}}
						}
			}}};
			\draw [thick,-{Latex[length=2mm]}] (-2.5,0) --(2.6,0) coordinate (xaxis);
			\draw [thick,-{Latex[length=2mm]}] (0.0,-2.1)coordinate(-yaxis) -- (0.0,1.0) coordinate (yaxis);
			\node [right] at (xaxis) {$\scriptstyle{\operatorname{Re}(z)}$};
			\node [right] at (yaxis) {$\scriptstyle{\operatorname{Im}(z)}$};
			\node  [below right] at (-yaxis) {\textcolor{darkred}{$\mathcal{C}^{\prime}$}};
			\draw[draw=darkred,fill=PastelRed, fill opacity=0.2,line width=0.8pt,rmidarrow] (-2.5,0.2) -- (2.5,0.2);  
			\begin{scope}
				\clip (-2.5,0.2) rectangle (-0.1,-2.7);   
				\draw[draw=darkred,fill=PastelRed,fill opacity=0.2, line width=0.8pt,dashed,midarrow] (0,0.2) circle(2.4);
			\end{scope}
			\draw[draw=red,line width=0.8pt,dashed,midarrow] (-0.1,-2.2)--(-0.1,0.1);
			\draw[draw=darkgreen,line width=0.8pt,dashed,midarrow] (0.1,0.05)--(0.1,-2.2);
			\fill[PastelRed, fill opacity=0.2] (-0.1,0.2)--(1.1,0.2)--(1.1,0.0)--(-0.1,0.0) ;
			\begin{scope}
				\clip (0.1,0.1) rectangle (1.0,-2.7);   
				\draw[draw=darkgreen,fill=PastelGreen, fill opacity=0.2, line width=0.8pt,dashed,midarrow] (0,0.2) circle(2.4);
			\end{scope}
			\draw[draw=darkgreen,line width=0.8pt,dashed,midarrow] (0.9,-1.8)--(0.9,0.1);
			\draw[draw=darkred,line width=0.8pt,dashed,midarrow] (1.1,0.1)--(1.1,-1.8);
			\begin{scope}
				\clip (1.0,0.2) rectangle (2.5,-2.7);   
				\draw[draw=darkred,fill=PastelRed, fill opacity=0.2, line width=0.8pt,dashed,midarrow] (0,0.2) circle(2.4);
			\end{scope}
			\draw[draw=yellow, thick] (0.0,0)--(1.0,0);
			\node [below left] {$O$};
			\draw[draw=darkgreen] (0.5,-0.5) circle(0.1);
			\node  at (0.5,-0.5) {\textcolor{darkgreen}{$\times$}};		
			\draw[draw=darkred] (-0.3,0.0) circle(0.1);
			\node  at (-0.3,0.0) {\textcolor{darkred}{$\times$}};		
			\draw[draw=darkred] (1.4,0.0) circle(0.1);
			\node  at (1.4,0.0) {\textcolor{darkred}{$\times$}};
		\end{tikzpicture}	
		\begin{minipage}{0.95\textwidth}		 
			\caption{Contour including the second Riemann sheet. At small coupling, the dominant contribution to the survival probability comes from the residue of the closest pole to the cut on the real axis. This residue produces an exponential decay intermediate asymptotic. For larger times,  the survival probability is dominated by  the line integrals parallel to the $\operatorname{Im}z$  axis. Their contribution determines the genuine asymptotic behavior consistently with Khalfin's theorem, see \eqref{sp-Lee:Riemann2}. \label{fig:contour2}}
		\end{minipage}
	\end{subfigure}
	\caption{Contours for the evaluation of the survival probability of the non-relativistic Lee model.
		\label{fig:contour}}
\end{figure}
Unitary evolution requires the Laplace transform 
\begin{align}
	 \hat{\alpha}(z)=\frac{1}{z-\omega+\omega\,\varkappa^{2}\,\tilde{\Lambda}(z)}
	\label{mqm-Lee:Lpa}
\end{align}
of the probability amplitude $\alpha(t)$  to be an analytic function on the entire complex plane except the real axis. In (\ref{mqm-Lee:Lpa}) $\tilde{\Lambda}$ denotes the non-dimensional level shift function. 

On the real axis, (\ref{mqm-Lee:Lpa}) can have poles and a cut,  corresponding to the discrete and continuous part of the energy spectrum, respectively. We obtain the level shift function on the full complex plane  
\begin{align}
	\tilde{\Lambda}(z)=\int_{-\Delta}^{\Delta}\mathrm{d}u\frac{1}{\xi+\omega-z}=\ln(\omega+\Delta-z)-\ln(\omega-\Delta-z)
\end{align}
by interpreting the logarithm of a complex number as the sum
\begin{align}
	\ln z=\ln|z|+\operatorname{Arg}(z)
	\label{mqm-Lee:log}
\end{align}
where the $\operatorname{Arg}$ function takes values in $[-\pi,\pi]$ and is defined by
\begin{align}
	\operatorname{Arg}(z):=-\pi\,\theta(-\operatorname{Re}z)\big{(}\theta(-\operatorname{Im}z)-\theta(\operatorname{Im}z)\big{)}+\arctan\frac{\operatorname{Im}z}{\operatorname{Re}z}
\end{align}
The upshot is that (\ref{mqm-Lee:Lpa}) has a cut in the interval $[\omega-\Delta,\omega+\Delta] $ on the real axis across which the level shift function has a discontinuity of $2\,\pi$ .

\begin{figure}[H]
	\begin{subfigure}{.5\textwidth}
		\centering
		\includegraphics{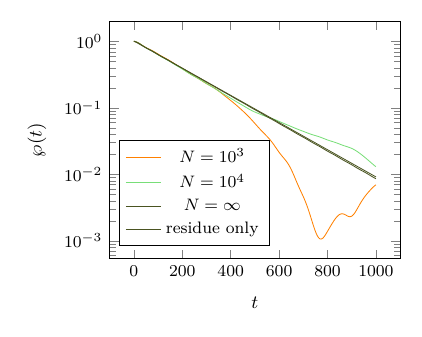}
		\caption{$\sigma=\sqrt{1.5\,\Gamma\, \Delta}$ as in Fig.~\ref{fig:Hexp} \label{fig:Lee1}}
	\end{subfigure}%
	\begin{subfigure}{.5\textwidth}
		\centering
		\includegraphics{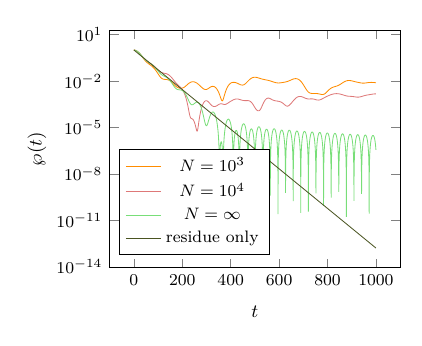}
		\caption{ $2.5 \,\sigma$. \label{fig:Lee2}}
	\end{subfigure}
	\begin{subfigure}{.5\textwidth}
		\centering
		\includegraphics{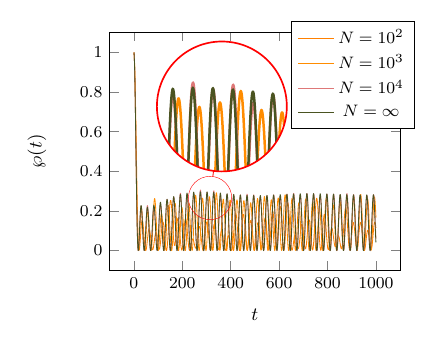}
		\caption{$7.5 \,\sigma$. \label{fig:Lee3}}
	\end{subfigure}%
	\begin{subfigure}{.5\textwidth}
		\centering
		\includegraphics{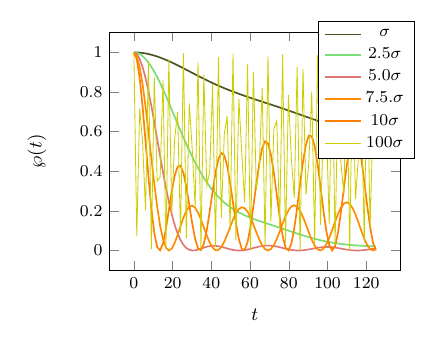}
		\caption{$N=\infty$ \label{fig:Lee4}}
	\end{subfigure}
	\caption{Evolution of the survival probability (ordinate) versus time (abscissa). We compute the  survival probability using one realization of single particle sector Hamiltonian $\operatorname{H}_{\scriptscriptstyle{1}}^{\scriptscriptstyle{(N)}}$  satisfying hypotheses \ref{H1}-\ref{H3}. We compare the result with the $N=\infty$ limit described by the non relativistic Lee model (\ref{mqm-Lee:sp-Lee}). Numerical values of parameters as in  Fig~\ref{fig:Hexp}. Figs~\ref{fig:Lee1}-\ref{fig:Lee2} \label{fig:Lee} are in semi-log scale to ease the comparison with the exponential decay intermediate asymptotics (\ref{mqm-Lee:vHove}) blown up by the weak coupling (van Hove) scaling limit. As the strength of the coupling increases \ref{fig:Lee3} the contribution of poles on the real axis becomes progressively more important and oscillations appear. In Fig~\ref{fig:Lee3}  we summarize the predictions of the non-relativistic Lee model for the survival probability of a local excitation as $\sigma$ varies. }
\end{figure}

The knowledge of (\ref{mqm-Lee:Lpa}) allows us to compute the survival probability by applying Cauchy's theorem to a line integral on the complex plane along the contour $\mathcal{C}$ of Fig~\ref{fig:contour1}. For small values of the coupling (\ref{mqm-Lee:coupling}) the result is dominated by the contribution of the line-integral along the boundary of a region of the complex plane encircling the cut. This integral also brings about an intermediate asymptotics characterized by exponential decay. Experimentally, at weak coupling, this intermediate asymptotic is likely to be the only measurable effect (Fig.~\ref{fig:Lee1}). To understand the origin of exponential decay, we can deform the contour of integration to $\mathcal{C}^{\prime}$ which has a section on the second Riemann sheet Fig~\ref{fig:contour2}. There $\hat{\alpha}_{\scriptscriptstyle{II}}$ has certainly poles outside the real axis. This is a model independent consequence of the analytic extension of a level shift function across a cut (see Appendix~\ref{app:Lee-poles}). The observed exponential decay is due to the residue of the pole closest to the real axis. The latter is a root for $y<0$ of the system
\begin{align}
	\label{mqm-Lee:pole}
	\begin{cases}
		x-\omega+\dfrac{\omega\,\varkappa^{2}}{2}\ln\dfrac{(\omega+\Delta-x)^{2}+y^{2}}{(\omega-\Delta-x)^{2}+y^{2}}=0
		\\[0.3cm]
			y+\omega\,\varkappa^{2}\left(\arctan\dfrac{y}{\omega+\Delta-x}-\arctan\frac{y}{\omega-\Delta-x}\right)+\pi\,\varkappa^{2}=0
		\end{cases}
	\end{align}
For small values of the coupling constant by solving (\ref{mqm-Lee:pole}) we obtain the intermediate asymptotics 
\begin{align}
&	\wp(t)\simeq e^{-2\,\pi\,\omega\,\varkappa^{2}\,t} &&\text{for} & \frac{1}{\operatorname{Var}_{\operatorname{P}_{\psi}}(\operatorname{H}_{\scriptscriptstyle{1}}^{\scriptscriptstyle{(sp)}})}\lesssim t \ll \infty
	\label{mqm-Lee:vHove}
\end{align}
where $\operatorname{Var}_{\operatorname{P}_{\psi}}(\operatorname{H}_{\scriptscriptstyle{1}}^{\scriptscriptstyle{(sp)}})$ is the variance of the energy over a local excitation. 
The lower bound on the interval is imposed by the {\textquotedblleft}Zeno regime{\textquotedblright} (see section~\ref{sec:Zeno}). The upper bound depends on the genuine time asymptotics of unitary evolution that is subject to Khalfin's theorem \cite{KhaL1957} (see also \cite{FoGhRi1978,PerA1980,GaVi1988,AnaC2018}). Khalfin's theorem, on its turn, is a consequence of Paley-Wiener theorem \cite[Theorem XII. ]{PaWi1934} and imposes for any unitary dynamics 
\begin{align}
	\int_{\mathbb{R}}\mathrm{d}t \frac{|\ln \wp(t)|}{1+t^{2}}<\infty
\end{align}
In writing this inequality, we have rescaled the time variable to a non-dimensional quantity.
Indeed, the genuine time asymptotics of the dynamics is determined by the line integrals along contour sections parallel to the imaginary axis in   Fig~\ref{fig:contour2}. In other words, (\ref{mqm-Lee:vHove}) corresponds to the prediction of van Hove's scaling limit which holds over time scales   
\begin{align}
	\pi\,\omega\,\varkappa^{2}\,t\sim O(1)
\end{align}
where the dynamics of the central qubit may be reliably described \cite{DavE1974} by a completely positive mater equation \cite{LinG1976,GoKoSu1976} see also \cite{ChrD2022} for a recent review and background on these concepts.

As the strength of the coupling increases, the contribution of the residues of the poles located on the real axis becomes increasingly important (Fig~\ref{fig:Lee3}).  These poles are the roots of the equation
\begin{align}
&	x-\omega+\frac{\omega\,\varkappa^{2}}{2}\ln\frac{(\omega+\Delta-x)^{2}}{(\omega-\Delta-x)^{2}}=0 && x \notin (\omega-\Delta,\omega+\Delta)
\end{align}
We can straightforwardly find these roots in the strong coupling expansion $\varkappa\uparrow\infty$
\begin{align}
	r_{\pm}=\pm \sqrt{2\,\omega\,\Delta}\,\varkappa
	+O(\varkappa^{0})
\end{align} 
Correspondingly, see Fig~\ref{fig:Lee4}, the dynamics of the survival probability becomes purely oscillatory with leading order contribution
\begin{align}
	\wp(t)\simeq 
	\cos^{2}\left(\sqrt{2\,\omega\,\Delta}\,\varkappa\,t
	\right)
\end{align}
Fig.~\ref{fig:Lee} provides quantitative illustrations of how the typical behavior of the survival probability varies as the strength of  the coupling (\ref{mqm-Lee:coupling}) increases. In particular, Figs~\ref{fig:Lee4} shows the existence of a parametric region at weak coupling where increasing the numerical value of the coupling makes exponential behavior steeper. This is possible if the contribution of the residues of the poles on real axis remains negligible in comparison to that on the second Riemann sheet. A further increase of the coupling constant produces, as expected, the appearance of oscillations at large times.

\subsection{Extension to a non diagonal environment}
\label{sec:mqm-ext}

Finally, in all our numerical experiments we see that the predictions of the non-relativistic Lee model appear in qualitative  agreement with the survival probability computed from a single particle Hamiltonian where $\Omega$ has non vanishing off-diagonal elements, with the same Gaussian distribution of hypothesis \ref{H3}. The following heuristic argument suggests an explanation for the numerical findings. 

As above, we suppose the single particle Hamiltonian to be real. In such a case, there exists an orthogonal transformation $\operatorname{T}$ that diagonalizes $\Omega$
\begin{align}
	\begin{bmatrix}
	1 & 0	\\  0 & \operatorname{T}
	\end{bmatrix}
	\begin{bmatrix}
		\omega & g^{\top} \\ g & \Omega
	\end{bmatrix}
	\begin{bmatrix}
		1 & 0	\\  0 & \operatorname{T}^{\top}
	\end{bmatrix}
	=
	\begin{bmatrix}
		\omega & g^{\top}\operatorname{T}^{\top} \\ \operatorname{T}g & \operatorname{diag}(\Omega)
	\end{bmatrix}
\end{align}  
As $\Omega$ is independent of $g$ so is $\operatorname{T}$. As a consequence, we can evaluate the Gaussian average in
(\ref{mqm-Lee:Ewp}) to obtain
\begin{align}
	F_{N}(z_{\mathfrak{1}},z_{\mathfrak{2}},s_{\mathfrak{1}},s_{\mathfrak{2}})=\operatorname{E}\sqrt{\det \operatorname{M}^{-1}}
	\nonumber
\end{align} 
where the expectation value is over the probability distribution of the entries of $\Omega$ and
\begin{align}
	\operatorname{M}_{i,j}=\delta_{i,j}+2\,\imath\,\sigma_{N}^{2}\,\sum_{\ell=1}^{N}\operatorname{T}_{\ell,i}\operatorname{T}_{\ell,j}
	\left(\frac{s_{\mathfrak{1}}}{z_{\mathfrak{1}}-\lambda_{\ell}(\Omega)}-\frac{s_{\mathfrak{2}}}{z_{\mathfrak{2}}-\lambda_{\ell}(\Omega)}\right)
\end{align}
We denote by $\lambda_{\ell}(\Omega)$ the $\ell$-th eigenvalue of $\Omega$.

The expectation value over the statistics of $\Omega$ is complicated by our physically motivated assumption that the statistical distribution of qubit level splittings, i.e. the diagonal elements of $\Omega$, is qualitatively different from that of qubit couplings, i.e. the off diagonal elements. Hence we cannot straightforwardly invoke the theory of orthogonal ensembles see e.g. \cite[\S~5]{PoBo2020}. Motivated by that theory, we can nevertheless make some assumptions, that seem plausible. To start with, we assume that in the limit of  $N$ tending to infinity, the eigenvalues and eigenvectors become statistically independent. Indeed, this is the case for orthogonal ensembles \cite[Corollary 2.5.4]{AnGuZe2009}. Second, we assume that the univariate marginal of the eigenvalue distribution is the same for all eigenvalues as it is the case for orthogonal ensembles.  Under these hypotheses, a straightforward calculation yields
\begin{align}
	\lim_{N\uparrow \infty}F_{N}(z_{\mathfrak{1}},z_{\mathfrak{2}},s_{\mathfrak{1}},s_{\mathfrak{2}})=\exp\left(\sigma^{2}\,\imath\operatorname{E}\left(\frac{s_{\mathfrak{1}}}{\xi-z_{\mathfrak{1}}}-\frac{s_{\mathfrak{2}}}{\xi -z_{\mathfrak{2}}}\right)\right)
\end{align}
 The expectation value on the random variable $\xi$ is again the Stieltjes transform of the univariate marginal of the $\Omega$-eigenvalue distribution. In the case of orthogonal ensembles, it is justified \cite[\S~5.3.2]{PoBo2020} to assume
that the Stieltjes transform has a single branch cut on the real axis. If this is true also in our case, then the qualitative behavior of the limit $N$ tending to infinity is indeed governed by a generalization of the non-relativistic Lee model, with continuous energy density eventually non-uniform.

The simplest explicit example is a Rosenzweig-Porter type model \cite{VeCuScTa2023} in which we posit
\begin{align}
	\Omega=\omega\operatorname{1}_{N-1}+\sigma_{N}\,\operatorname{G}
	\label{mqm-ext:RP}
\end{align}
where $\operatorname{G}$ belongs to the Gaussian orthogonal ensemble. In such a case, we get
\begin{align}
	\xi\overset{\scriptscriptstyle{law}}{=}\omega+	\eta
\end{align}
where $\eta$ follows the Wigner distribution. Accordingly, the level shift function is the Stieltjes transform of the Wigner distribution:
\begin{align}
	\operatorname{E}\frac{1}{\eta+\omega-z}=\int_{-2\sigma}^{2\sigma}\frac{\mathrm{d}w}{\sigma\,\pi}\frac{1}{w+\omega-z}\sqrt{1-\frac{w^{2}}{4\,\sigma^{2}}}=-\frac{(z-\omega)-\sqrt{(z-\omega)^{2}-4\,\sigma^{2}}}{2\,\sigma^{2}}	
\end{align}
Upon inserting this expression into the survival probability, it is straightforward to verify that the survival probability is exactly given by (\ref{mqm-Lee:Lpa}). We have thus obtained an interesting result. The relaxation properties of a local excitation in an analytically solvable chain, with only nearest neighbor interactions are exactly the same of those of a local excitation of a multi-qubit system whose interactions are described by the Gaussian orthogonal ensemble. A long standing conjecture \cite{BoGiSc1984} upholds that {\textquotedblleft}spectra of time-reversal-invariant systems whose classical analogs are K-systems (this means, roughly speaking, strongly mixing measure preserving) show the same fluctuation properties as predicted by the Gaussian orthogonal ensemble{\textquotedblright}, see \cite{HaGnKu2018} for a recent account on quantum chaos theory. Thus, both an {\textquotedblleft}integrable{\textquotedblright} and a {\textquotedblleft}chaotic{\textquotedblright} model yield the same expression for the survival probability.
We emphasize that both models share the property of having eigenvectors, or equivalently, normal modes, with delocalized components.
Our interpretation is that this phenomenon is a further manifestation of a general mechanism producing equilibration in classical and quantum statistical mechanics \cite{CaBaLuMGVu2025}: the dependence of the observable, in our case the survival probability of a local excitation, on a large number of extended normal modes of the microscopic dynamics. 
  
\section{Speed limits on decorrelation}
\label{sec:Zeno}

\begin{figure}
	\centering
		\includegraphics[width=0.7\textwidth]{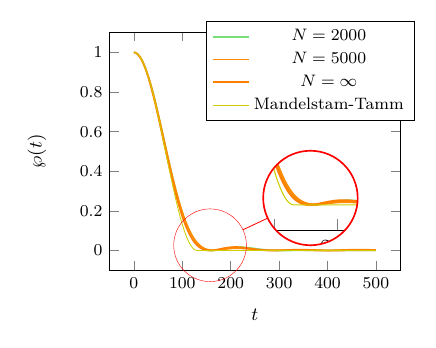}
	\caption{Evolution of the survival probability (ordinate) versus time (abscissa) for one realization of  $\operatorname{H}_{\scriptscriptstyle{1}}^{\scriptscriptstyle{(N)}}$ sampled from the Rosenzweig-Porter type model (\ref{mqm-ext:RP}) with same parameters $\omega$, $\sigma$ as in Fig~\ref{fig:Hexp}. For large $N$  the survival probability tends to (\ref{sp-explicit:Bessel}). In the figure for times larger than (\ref{Zeno:splim}) we set the Mandelstam-Tamm lower bound to zero. \label{fig:goe}}
\end{figure}

The short time behavior of the survival probability is subject to a universal lower bound determined by the Mandelstam-Tamm uncertainty relation \cite{MaTa1945}. Namely, Fleming \cite{FleG1973}, albeit by a different route, and later Bhattacharyya \cite{BhaK1983} showed that for any state with finite variance
\begin{align}
	&	\wp(t)\,\geq\,\cos^{2}\left(\sqrt{\operatorname{Var}_{\operatorname{P}_{\psi}}(\operatorname{H})}\,t\right) && t\in \left [0,\frac{\pi}{2\,\sqrt{\operatorname{Var}_{\operatorname{P}_{\psi}}(\operatorname{H})}}\right ] 
	\label{Zeno:lb}
\end{align}
where we used the obvious definition
\begin{align}
	\operatorname{Var}_{\operatorname{P}_{\psi}}(\operatorname{H})=\left \langle\,\psi\,\,,\operatorname{H}^2\psi\,\right\rangle-\left(\left \langle\,\psi\,\,,\operatorname{H}\psi\,\right\rangle\right)^{2}
\end{align}
The lower bound also exists in the case of dynamics generated by  time dependent Hamilton operators \cite{PfFr1995} and can be extended to cases where the variance is infinite using a weaker formulation of the uncertainty relations \cite{UffJ1993}.
The upshot is that the time it takes to the survival probability of a pure state to significantly decrease from the unity cannot be shorter than
\begin{align}
	\tau\,=\,\frac{\pi}{2\,\sqrt{\operatorname{Var}_{\operatorname{P}_{\psi}}(\operatorname{H})}}
	\label{Zeno:splim}
\end{align} 
For our purposes, it is interesting to observe that the lower bound is saturated when the level shift function does not bring about a pole in the second Riemann sheet. In that case, no exponential decay region appears.   In Fig~\ref{fig:goe} we show the rapid convergence of the survival probability for the Rosenzweig-Porter type model (\ref{mqm-ext:RP}) to the Bessel function (\ref{sp-explicit:Bessel}) and, as already verified for the {\textquotedblleft}integrable{\textquotedblright} chain (\ref{Fock-exe1:H}) (see Fig~\ref{fig:explicit2}), the overlap of all these curves with the lower bound prediction (\ref{Zeno:lb}) for times shorter than (\ref{Zeno:splim}). In experimentally realizable situations, we expect an exponential decay region to generically appear.  
\begin{figure}
	\begin{subfigure}{.5\textwidth}
		\centering
		\includegraphics[width=\textwidth]{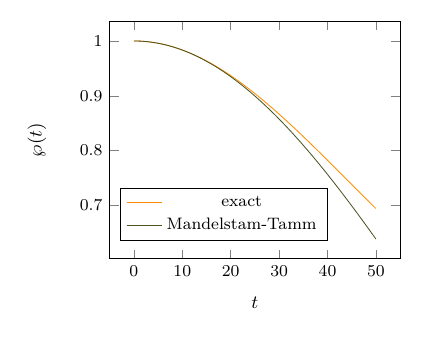}
		\caption{Survival probability for $N=10$\label{fig:MT1}}
	\end{subfigure}
	\begin{subfigure}{.5\textwidth}
		\centering
		\includegraphics[width=\textwidth]{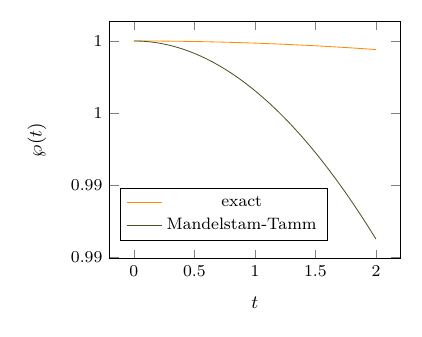}
		\caption{Survival probability for $N=10^{4}$ \label{fig:MT2}}
	\end{subfigure}
	\caption{Evolution of the survival probability (ordinate) versus time (abscissa). We compare one realization of  $\operatorname{H}_{\scriptscriptstyle{1}}^{\scriptscriptstyle{(N)}}$ sampled from the same ensemble used for Fig~\ref{fig:Hexp} with the Mandelstam-Tamm lower bound. The region where the lower bound  (\ref{Zeno:lb}) produced by the Mandelstam-Tamm uncertainty relation contracts to a limit value. When $\Omega$ is diagonal, the limit value is specified by the non-relativistic Lee model. \label{fig:MT}}
\end{figure}
In Fig~\ref{fig:MT} we show that for the realistic model (\ref{Fock-exe2:H}) that lower bound is close to the actual behavior only in a time interval that shrinks to a limit value as the number of qubits in the system increases. When hypotheses \ref{H1}-\ref{H3} 
apply the limit value is again governed by the non-relativistic Lee model.

\section{Extension of the Kac-Mazur-Montroll estimate to the quantum case}
\label{sec:Kac}

A second universal result controls the expectation value of the mean recurrence frequency to a value $p$ of the survival probability of a quantum system with a finite number of states $N$. The result is an extension to the quantum case of Mark Kac's  estimate \cite{KacM1943} upholding Boltzmann's reply to Zermelo's Wiederkehreinwand. To extend Kac's estimate to the quantum survival probability, we follow the approach of \cite{MaMo1960}. The technical difference is the use of a Hubbard-Stratonovich transformation to deal with the square of trigonometric polynomials.    Our aim is to prove that the mean frequency of recurrence of the survival probability to a value $p$ as $N$ tends to infinity grows exponentially with the number of states $N$
\begin{align}
&	\nu(p)\asymp e^{-\frac{N\,p}{\varkappa_{\star}}} && N\gg 1
\end{align}
Below, we specify the value of the exponent $ \varkappa_{\star}$ and replace the logarithmic asymptotics $ \asymp$ with an estimation of the prefactor. We emphasize here that the result crucially depends on the hypothesis that the local excitation has a finite projection on all, or at least a significantly large fraction, of the eigenstates of the system.  In addition, we assume eigenvectors to be delocalized. In other words, we hypothesize that in a  $N$ -dimensional system the generic component of any eigenvector is of magnitude $O(N^{-1/2})$. In random matrix theory, this is a property enjoyed by ensembles with independent entries \cite{RuVe2015}. All examples we consider satisfy this hypothesis.

We now turn to the proof of the claim in a few steps.

\subsubsection{Integral representation of the mean return frequency}

Following \cite{KacM1943}, we consider the function 
\begin{align}
	L(t,p)=\wp(t)-p
	\label{Kac:zeroes}
\end{align}
For fixed $p$, the function has a root each time the survival probability $\wp(t)$ returns to the value $p$. We write the number of these roots as
\begin{align}
	N(T,p)=\int_{-T}^{T}\mathrm{d}t\,\sum_{i: L(t_{i},p)=0}\delta(t-t_{i})
\end{align}
Correspondingly, we adopt Kac's definition of the mean recurrence frequency
\begin{align}
	\nu(p)=\lim_{T\uparrow \infty}\frac{1}{2\,T}N(T,p)
\end{align}
The idea of Mazur and Montroll in \cite{MaMo1960} for evaluating the mean frequency is to work with distributions. First, they take advantage of the properties of the Dirac's distribution to eliminate the sum over the roots:
\begin{align}
	N(T,p)=\int_{-T}^{T}\mathrm{d}t\,|(\partial_{t}L)(t,p)|\delta(L(t,p))
\end{align}
Next, they use the remarkable identity
\begin{align}
	|x|=\frac{1}{\pi}\int_{\mathbb{R}}\mathrm{d}y\frac{1-\cos(y\,x)}{y^{2}}
\end{align}
to obtain an integral representation of the absolute value. Using the Fourier transform of the Dirac's $\delta$-distribution, we arrive at
\begin{align}
	\nu(p)=\lim_{T\uparrow \infty}\frac{1}{T}\int_{\mathbb{R}_{+}^{2}}\frac{\mathrm{d}y\mathrm{d}z}{2\,\pi^{2}}\int_{-T}^{T}\mathrm{d}t\,\frac{1-\cos\left(y\,\dot{\wp}(t)\right)}{y^{2}}\,\cos\left(z\,\wp(t)-z\,p\right )
	\label{Kac:nu}
\end{align}
This representation of the number of roots is convenient because it allows us to apply the Kroenecker-Weyl phase averaging theorem, \cite{WeyH1916,WeyH1938} and Appendix~\ref{app:KW}, to exactly evaluate the time average and then the stationary point method to evaluate the remaining integrals.

\subsubsection{Application of the Kroenecker-Weyl phase averaging theorem}
 
We start by recasting the general expression of the survival probability (\ref{sp:sp}) into the form 
\begin{align}
	\wp(t)=\sum_{\ell=1}^{M} \cos (\epsilon_{\ell}t)c_{\scriptscriptstyle{\ell}}\sum_{k=1}^{N}\cos (\epsilon_{k}t)c_{\scriptscriptstyle{k}}+\sum_{\ell=1}^{N} \sin (\epsilon_{\ell}t)\,c_{\scriptscriptstyle{\ell}}\sum_{k=1}^{N}\sin (\epsilon_{k}t)\,c_{\scriptscriptstyle{k}}
\end{align}
We recall that the coefficients $c_{\scriptscriptstyle{k}}$ denote the probability to measure the system in the $k$-th energy state after having initially prepared it in a state describing a local excitation. By definition, we get
\begin{align}
	\sum_{k=1}^{N}c_{\scriptscriptstyle{k}}=1
\end{align}
Correspondingly, we get
\begin{align}
	\dot{\wp}(t)&=-2\sum_{\ell=1}^{N} \cos (\epsilon_{\ell}t)c_{\scriptscriptstyle{\ell}}\sum_{k=1}^{N}\epsilon_{k}\sin (\epsilon_{k}t)c_{\scriptscriptstyle{k}}+2\sum_{\ell=1}^{N} \sin (\epsilon_{\ell}t)\,c_{\scriptscriptstyle{\ell}}\sum_{k=1}^{N}\epsilon_{k}\cos (\epsilon_{k}t)\,c_{\scriptscriptstyle{k}}
\end{align}
Kroenecker-Weyl's phase averaging theorem  allows us to replace the time average over these trigonometric sums with the average over their phases. Specifically, we can write (\ref{Kac:nu}) into the form of a multiple integral over the phase n-tuple $\phi=[\phi_{\scriptscriptstyle{1}},\dots,\phi_{\scriptscriptstyle{n}}] $
\begin{align}
	\nu(b)=2\operatorname{Re}\int_{\mathbb{R}_{+}^{2}}\frac{\mathrm{d}z\mathrm{d}y}{2\,\pi^{2}} 
	\prod_{ j=1}^{N}\int_{0}^{2\pi}\frac{\mathrm{d}\phi_{j}}{2\,\pi} \,e^{\imath z V(\bm{\phi})-\imath\,z\,p}
	\frac{2- e^{\imath\,y\,W(\bm{\phi})}-e^{ -\imath\,y\,W(\bm{\phi})}}{2\,y^{2}}
	\label{Kac:start}
\end{align}
The integrand depends on
\begin{align}
	&	V(\bm{\phi})=\sum_{\ell=1}^{N} \cos \phi_{\ell}\,c_{\scriptscriptstyle{\ell}}\sum_{k=1}^{N}\cos \phi_{k}\,c_{\scriptscriptstyle{k}}+\sum_{\ell=1}^{N} \sin \phi_{\ell}\,c_{\scriptscriptstyle{\ell}}\sum_{k=1}^{N}\sin \phi_{k}\,c_{\scriptscriptstyle{k}}
\\
	&W(\bm{\phi})=-2\sum_{\ell=1}^{N} \cos \phi_{\ell}\,c_{\scriptscriptstyle{\ell}}\sum_{k=1}^{N}\epsilon_{k}\sin \phi_{k}\,c_{\scriptscriptstyle{k}}+2\sum_{\ell=1}^{N} \sin \phi_{\ell}\,c_{\scriptscriptstyle{\ell}}\sum_{k=1}^{N}\epsilon_{k}\cos \phi_{k}\,c_{\scriptscriptstyle{k}}
\end{align}
These quantities differ from those encountered in classical statistical mechanics by \cite{KacM1943} and \cite{MaMo1960} in as much they are quadratic in the trigonometric functions. This fact apparently prevents us from exactly performing the phase averages using 
\begin{align}
	\mathscr{J}_{0}\left(\sqrt{x^{2}+y^{2}}\right)=\int_{0}^{2\pi}\frac{\mathrm{d}\phi}{2\,\pi}e^{\imath\,x \cos\phi+\imath\,y\sin \phi}
	\label{Kac:Bessel}
\end{align}
where $\mathscr{J}_{0}$ is Bessel function of the first kind of order zero as done in \cite{KacM1943,MaMo1960}. 
We can circumvent this stumbling block by resorting to a complex  Hubbard-Stratonovich transform.  We refer to Appendix~\ref{app:Integrals} for details of the calculation. After applying the transform, the arguments of the exponentials become linear in the trigonometric functions at the price of introducing extra integrations over auxiliary variables.
We are now in a position to perform the phase averages using (\ref{Kac:Bessel}).  The upshot is 
\begin{align}
			\nu(b)=2\,\operatorname{Re}\imath\int_{\mathbb{R}_{+}^{2}}\frac{\mathrm{d}z\mathrm{d}y}{2\,\pi^{2}}
		\,e^{-\imath\,p\,z} \int_{\mathbb{R}^{2}}\frac{\mathrm{d}a_{1}\mathrm{d}a_{2}}{\pi}\int_{\mathbb{R}^{4}}\frac{\prod_{i=1}^{4}\mathrm{d}x_{i}}{\pi^{2}}\,I(z,y,a_{1},a_{2},x_{1},\dots,x_{4})
			\label{Kac:pares1}
		\end{align}
		where
	\begin{align}
	&	I(z,y,a_{1},a_{2},x_{1},\dots,x_{4})=z\,
		e^{-4\,\imath\,z\,(a_{1}^{2}+a_{2}^{2})-\imath\,y\, (x_{1}x_{2}+x_{3}x_{4})}
		\prod_{j=1}^{N} \left(\mathscr{J}_{0}\left(4\,c_{\scriptscriptstyle{j}}\, z\sqrt{a_{\mathfrak{1}}^{2}+a_{\mathfrak{2}}^{2}}\right)\right .
	\nonumber\\&\hspace{1.0cm}	\left .	-\mathscr{J}_{0}\left(c_{\scriptscriptstyle{j}} \sqrt{\big{(}4 \,z\,a_{1}-y\,x_{1}+2\,\epsilon_{k}\,y\,x_{4}\big{)}^{2}+\big{(}4 \,z\,a_{2}-y\,x_{3}-2\,\epsilon_{k}\,y\,x_{2}\big{)}^{2}}\right)\right)
	\label{Kac:pares2}
\end{align}

\subsubsection{Use of the delocalization hypothesis}. We are interested in the limit $N$ tending to infinity. Using probability conservation and the fact that eigenvector components are typically of magnitude $O(N^{-1/2})$ , we set
\begin{align}
	&	\varkappa:=\sum_{i=1}^{N}c_{k}^{2}\simeq \frac{\varkappa_{\star}}{N}
\\
	&\Gamma:=\sum_{i=1}^{N}\,c_{k}^{2}\,\epsilon_{k}^{2}\simeq \frac{\Gamma_{\star}}{N}
\end{align}
and
\begin{align}
	\gamma:=\sum_{i=1}^{N}\,c_{k}^{2}\,\epsilon_{k}\simeq \frac{\gamma_{\star}}{N}
\end{align} 
These definitions imply
\begin{align}
	\varkappa\,\Gamma-\gamma^{2}\,\geq\,0
\end{align}

\subsubsection{Integration over the Hubbard-Stratonovich auxiliary variables}. 
We avail us  of the same observation made in \cite{KacM1943,MaMo1960} that as $N$ tends to infinity the integral concentrates near the origin. A saddle point like approximation then boils down to evaluate the integral by setting
\begin{align}
	\prod_{j=1}^{N} \mathscr{J}_{0}\left(4\,c_{\scriptscriptstyle{j}} z\sqrt{a_{\mathfrak{1}}^{2}+a_{\mathfrak{2}}^{2}}\right)
	\simeq e^{-4\,\varkappa\,z^{2}\,(a_{\mathfrak{1}}^{2}+a_{\mathfrak{2}}^{2})}
\end{align}
and
\begin{align}
	& 	\prod_{j=1}^{N} \mathscr{J}_{0}\left(c_{\scriptscriptstyle{j}} \sqrt{\big{(}4 \,z\,a_{1}-y\,x_{1}+2\,\epsilon_{k}\,y\,x_{4}\big{)}^{2}+\big{(}4 \,z\,a_{2}-y\,x_{3}-2\,\epsilon_{k}\,y\,x_{2}\big{)}^{2}}\right)
	\nonumber\\
	& \simeq
	\exp\left(-4\, \varkappa \, z^2\left(a_1^2+a_2^2\right) -y^2 \left(\varkappa \,\frac{ x_1^2+ x_3^2}{4}+\Omega(x_2^2 +x_4^2)  -\gamma  \,(x_4
	x_1-  x_2 x_3)\right)\right)
	\nonumber\\
	&	\,\times\,\exp\left(2\, y\, z\, \Big{(}a_2 \,\left(2\, \gamma\,  x_2+\varkappa \, x_3\right)+a_1 \left(\varkappa \, x_1-2 \gamma \, x_4\right)\Big{)}\right)
\end{align}
The conceptually straightforward recursive evaluation of the resulting Gaussian integrals leaves us with the expression  
\begin{align}
	\nu(p)\simeq \,\operatorname{Re}\int_{\mathbb{R}_{+}^{2}}\frac{\mathrm{d}z\mathrm{d}y}{2\,\pi^{2}}\,
	\frac{e^{-\imath\,z\,p}}{y^{2}} \left(\frac{\imath }{\imath+\varkappa\,z}
	-\frac{\imath }{(1+y^{2}\,(\varkappa\,\Gamma-\gamma^{2}))\left(\imath+\frac{\varkappa\,z}{1+y^{2}\,(\varkappa\,\Gamma-\gamma^{2})}\right)}\right)
\end{align}
\subsubsection{Last step}
The real part of the first addend in the integrand is
\begin{align}
	\operatorname{Re}\left(\frac{\imath \,e^{-\imath\,z\,p}}{\imath+\varkappa\,z}\right)
	=\frac{\cos(z\,p)+\varkappa\,z\,\sin(z\,p)}{1+\varkappa^{2}\,z^{2}}
\end{align}
Applying Cauchy's residue theorem
\begin{align}
	& 	\int_{\mathbb{R}_{+}}\mathrm{d}z\left(\varkappa\,z\frac{\sin(z\,p)}{\left(\varkappa\,z\right)^{2}+1}+\frac{\cos(z\,p)}{\left(\varkappa\,z\right)^{2}+1}\right)=\frac{\pi}{\varkappa}\, e^{-\frac{p}{\varkappa}}
\end{align}
The second addend has the same structure, we therefore get
\begin{align}
	\nu(p)
	\simeq  \frac{e^{-\frac{p}{\varkappa}}}{4\,\pi\,\varkappa}\,\int_{\mathbb{R}}\mathrm{d}y\,\frac{1-e^{-p\,y^{2}\,\left(\Gamma-\frac{\gamma^{2}}{\varkappa}\right)}}{y^{2}}
\end{align}
This last integral is of the same form as the one found in the last step of the calculation in \cite{MaMo1960}. 
We can evaluate it exactly. We obtain:
\begin{align}
	\nu(p)\simeq \frac{\sqrt{p\,\left(\Gamma-\frac{\gamma^{2}}{\varkappa}\right)\,\pi}}{2\,\pi\,\varkappa}\,e^{-\frac{p}{\varkappa}}\simeq \frac{N^{1/2}}{2\,\varkappa_{\star}}\sqrt{\frac{p}{\pi} \,\left(\Gamma_{\star}-\frac{\gamma_{\star}^{2}}{\varkappa_{\star}}\right)}e^{-\frac{N\,p}{\varkappa_{\star}}}
\end{align}
The frequency is dimensionally the inverse of the mean return time
\begin{align}
&	\tau(p)\simeq \frac{2\,\varkappa_{\star}\,e^{\frac{N\,p}{\varkappa_{\star}}}}{N^{1/2}\sqrt{\frac{p}{\pi} \,\left(\Gamma_{\star}-\frac{\gamma_{\star}^{2}}{\varkappa_{\star}}\right)}} && N\gg 1
\label{Kac:revival}
\end{align}
We have therefore extended Kac's result to quantum systems with $N$ states. We emphasize that our derivation makes use of the general, non-perturbative, expression of the survival probability (\ref{sp:sp})  rather than the first order expression in perturbation theory used in \cite{KaWuClPe2026}.

\subsection{Numerical predictions}

Two aspects of the above calculation should be emphasized. First, the  Kac-Mazur-Montroll estimate is the result of an application of the steepest descent method. Hence, its prediction of exponential increase of the revival time with the size of the system $N$ is expected to be accurate when $N$ is large. The merit of the prediction is to hold for a large class of models of many-body systems. Second, the estimate, strictly speaking, applies to the average revival \emph{frequency}. While we can use the inverse of the the mean return frequency to estimate the order of magnitude of the average return time, as it was originally done in \cite{MaMo1960}, this is different from the estimate obtained in \cite{KaWuClPe2026} which applies directly to the mean return time but relies on a model dependent expression of the survival probability.  

\begin{figure}
	\centering
	\includegraphics[width=0.7\textwidth]{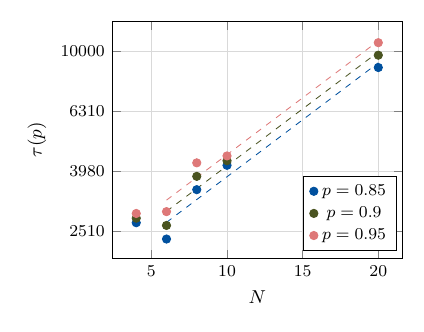}
	\caption{Return time versus size $N$ of the single particle sector in units of the central qubit frequency for the model considered in \ref{sec:mqm-Lee}. The color of the points correspond to distinct values of the return probability $p$. The values are obtained sampling a realization of the parameter according to the same distribution as in Fig~\ref{fig:Lee1}. The outlier at $N=4$ is an indication that applying the steepest descent method for $N=4$ is, as to be expected not very accurate. The size of the single particle sector is chosen for comparison with Figure~4 of \cite{KaWuClPe2026}. \label{KMM_revival}}
\end{figure}

Bearing these considerations in mind, we can apply the Kac-Mazur-Montroll estimate to the coupling model described in section~\ref{sec:mqm-Lee} using the same parameters as in Fig~\ref{fig:Lee1} and for small values of $N$. The scope is to contrast the order of magnitude of the revival time predicted by (\ref{Kac:revival}) with the results predictions of \cite{KaWuClPe2026}.
The upshot is that the predictions agree in order of magnitude, which is, based on the considerations at the beginning of this paragraph, what can be realistically expected.  It is fair to add that by repeating experiments, we verified that revival times for all values of $N$ reported in Fig~\ref{KMM_revival} remain sensitive to fluctuations in the value of the prefactor of the exponential in (\ref{Kac:revival}). This means, not surprisingly, that the values of $N$ are to small to produce self-averaging of the prefactor.
Said so, the information conveyed by the order of magnitude of the revival time is valuable.
The typical energy separation of qubits in \cite{WuYaAnAnCle2024} is of the order of $10\text{GHz}$. This means that translating into seconds the revival times in Fig~\ref{KMM_revival}, we obtain $O(\tau(p))\sim \SIrange{0.1}{1}{\micro\second}$ which is sufficiently separated from the millisecond time-scale of relaxation and dephasing  of state-of-the-art qubits \cite{KjScBrKrOl2020,BaMuShGr2024,TuSuGoVeMo2025}. We refer to \cite{KaWuClPe2026} for further discussion of this point.

\section{Conclusion and outlook}

A key motivation of the present study is the active development of multi-qubit platforms based on superconducting circuits \cite{AnAbMi2025,WuYaAnAnCle2024}. These systems offer a high degree of controllability: both the qubit frequencies and their mutual couplings can be precisely engineered, while their quantum states can be initialized, manipulated, and measured using well-established techniques. Experiments can therefore be performed either on tailor-made circuits, or on small-scale quantum processors. An important feature of the present-day superconducting qubits is that their relaxation and dephasing times are approaching a millisecond \cite{KjScBrKrOl2020,BaMuShGr2024,TuSuGoVeMo2025}, which means that the interesting dynamics of it happens on time scales much shorter than this. The latter scale is determined by the typical frequency of the qubits, generally in 10 GHz regime. This pronounced separation in the time scales, millisecond versus nanosecond enables the experimental investigation of nearly unitary evolution, as explored in this work.

In the present work, we focus on the survival probability for the one-particle sector. Under rather general assumptions, we showed that the model is exactly solvable when the univariate marginal of the eigenvalue distribution is known. An important physical requirement on the random matrix ensemble specifying the Hamiltonian is that on-diagonal entries must follow a probability distribution descriptive of realizable qubit level splittings. This distribution is necessarily markedly different 
from the one that describes coupling constants.  The results of \cite{VeCuScTa2023} on the Rosenzweig-Porter model provide a promising way to obtain explicit, though approximate expression for the eigenvalue distribution in cases of experimental interest. 

From the theoretical point of view a systematic analysis of the survival probability in all the sectors of the multi-qubit system is a question that deserves further exploration. We expect the survival probability in sectors with few particle excitations to exhibit a qualitative behavior similar to that of the one-particle sector. As the number of excitations increases, however, the situation should change: we readily see that the $N$ particle sector cannot exhibit any decay.  Further interesting questions concern how the properties of the survival probability are affected by those of the random matrix ensemble used to model the Hamilton operator generating the $n$-particle sector dynamics. Physically, the realization of qubits requires a mesoscopic quantum system confined by a non-harmonic potential. This is necessary to effectively reduce the dynamics to that of a two-level system. In the present work, we neglect anharmonic effects. In their presence, distinct $n$-particle sectors are generically coupled. In a first approximation, we can imagine to treat this coupling as an environmental effect that can be described by embedding the unitary dynamics of a $n$-particle sector, conceptualized as an open system, in a quantum master equation of Lindblad form, see e.g. \cite{BrPe2002}.   

In this work, we restrict the attention on qubit chains. The Kac-Mazur-Montroll estimate indicates that the relaxation of the survival probability, in agreement with a generic mechanism providing for equilibration and thermalization (see e.g. \cite{CaBaLuMGVu2025}), occurs because the initial state has a projection on a large fraction of the eigenstates of the Hamilton operators, which are often expected to be delocalized for  many-body systems, see, however, \cite{AbPa2017} for important counter-examples. The topology of the interactions can impinge on this picture. Thus, extensions of our model to be considered involve the study of the qubit survival probability on graph-like architectures,  such those corresponding to a star or a tree topology of the platform. Increasing the complexity of the topology complicates the analytical treatment. Interestingly, recent progress  indicates that integrable or semi-integrable models can be nevertheless recovered for quantum many-body systems distributed on loop-less lattices \cite{PaPaPiRoSc2019,AuMuTu2022}.

We leave these questions for future theoretical work. 

\section{Acknowledgments}

P.M-G warmly thanks L. Magazz\'u, for discussions, careful reading of the manuscript, and many useful comments, and M. Cattaneo for many discussions, and for pointing out to us refs \cite{RuVe2015,VeCuScTa2023,KoOs2026}.  P.M-G acknowledges a partial leave from teaching duties supported by the Centre of Excellence FiRST of the Research Council of Finland. The work of B.K. is funded by the European Union’s Research and Innovation Programme, Horizon Europe, under the Marie Sklodowska-Curie Grant Agreement No.
101150440 (TcQTD). B.K. and J.P. acknowledge QuantERA II Programme that has received funding from the EU’s H2020 research and innovation programme under the GA No 101017733, and the Research Council of Finland Centre of Excellence programme grant 336810 and grant 349601 (THEPOW).
\vskip7ex
\begin{center}
	* * 
\end{center}

\section{Statements and Declarations: Conflict of interest and Data availability}

On behalf of all authors, the corresponding author states that there is no
conflict of interest and data sharing is not applicable to this article as no datasets were generated or analyzed during the current study.

\appendix
\counterwithin*{equation}{section}
\renewcommand\theequation{\thesection\arabic{equation}}

\section*{Appendices}

\section{Linear algebra identities used for the expression of the resolvent}
\label{app:LA}

We recall without proofs some results in linear algebra that we use in the main text. We refer ,e.g., to \cite[\S~0.7.3, \S~0.7.4]{HoJo2013} or to the Wikipedia articles \cite{Wiki:BM,Wiki:SM} for proofs and further analysis. 

Let $\operatorname{M}$ be a non singular square matrix, admitting the block decomposition
\begin{align}
	\operatorname{M}=
	\begin{bmatrix}
	\operatorname{M}_{1,1} & 	\operatorname{M}_{1,2}	\\  \operatorname{M}_{2,1} & 	\operatorname{M}_{2,2}	
	\end{bmatrix}
\end{align}
We also suppose that the diagonal blocks  $\operatorname{M}_{1,1}$, $\operatorname{M}_{2,2}$ are also non singular square matrices. Then the inverse of $\operatorname{M}$ is well defined and  
admits the representation
	\begin{align}
	\operatorname{M}^{-1}
	=
	\begin{bmatrix}
		\operatorname{G}_{1,1}^{-1} &- \operatorname{G}_{1,1}\operatorname{M}_{12}\operatorname{M}_{22}^{-1}	\\  -\operatorname{G}_{2,2}^{-1}\operatorname{M}_{21}\operatorname{M}_{11}^{-1} &\operatorname{G}_{2,2}^{-1}
	\end{bmatrix}
	\label{LA:bi}
\end{align}
with
\begin{align}
&	\operatorname{G}_{1,1}=\operatorname{M}_{1,1} -\operatorname{M}_{1,2}\operatorname{M}_{2,2}^{-1}\operatorname{M}_{2,1}
\\
& \operatorname{G}_{2,2}=\operatorname{M}_{2,2} -\operatorname{M}_{2,1}\operatorname{M}_{1,1}^{-1}\operatorname{M}_{1,2}
\end{align}
In writing, (\ref{LA:bi}) we expressed off diagonal elements using the identities
\begin{align}
&	\operatorname{G}_{1,1}\operatorname{M}_{1,2}\operatorname{M}_{2,2}^{-1}=\operatorname{M}_{1,1}^{-1}\operatorname{M}_{1,2}\operatorname{G}_{2,2}
\\
&\operatorname{G}_{2,2}\operatorname{M}_{2,2}\operatorname{M}_{1,1}^{-1}=\operatorname{M}_{2,2}^{-1}\operatorname{M}_{2,1}\operatorname{G}_{1,1}
\end{align}
that can be proven using e.g. by means the Taylor series that define $\operatorname{G}_{1,1}$ and $\operatorname{G}_{2,2}$.

In order to derive the expression of the lower diagonal  blocks of the resolvent in (\ref{mqm:res}), we
also avail us of Sherman-Morrison formula.  This latter identity relates the inverse of the sum of a non-singular matrix $\operatorname{M}$ with the one dimensional projector along the span of a vector $v$ to the inverse of $\operatorname{M}$ alone:
\begin{align}
	\left(\operatorname{M}+v\,v^{\dagger} \right)^{-1}
	=\operatorname{M}^{-1}+\frac{\operatorname{M}^{-1}\,vv^{\dagger}\operatorname{M}^{-1}}{1+\left \langle\,v\,,\operatorname{M}^{-1}v\,\right\rangle}
	\label{LA:SM}
\end{align}

\section{Expression of the survival probability at sub-leading order in perturbation theory}
\label{app:hopt}

After straightforward algebra, that can be efficiently implemented also by software for analytic manipulations, we arrive at the expression
\begin{align}
		\wp(t)=&1
	-4 \,\varepsilon^{2}\sum_{j\neq 1}\sin^{2}\left(\frac{\epsilon_{1,j}^{\scriptscriptstyle{(0)}}t}{2}\right)\,\frac{|\operatorname{V}_{1,j}^{\scriptscriptstyle{(N)}}|^{2}}{\epsilon_{1,j}^{\scriptscriptstyle{(0)}}{}^{2}}\left(1-\varepsilon^{2}\sum_{k\neq 1} \frac{\left |\operatorname{V}_{1,k}^{\scriptscriptstyle{(N)}}\right |^{2}}{\epsilon_{1,k}^{\scriptscriptstyle{(0)}}{}^{2}}-\varepsilon^{2}\sum_{k\neq j} \frac{\left |\operatorname{V}_{j,k}^{\scriptscriptstyle{(N)}}\right |^{2}}{\epsilon_{j,k}^{\scriptscriptstyle{(0)}}{}^{2}}\right)
	\nonumber\\
	&-8\,\varepsilon^{3}\sum_{j\neq 1}\sin^{2}\left(\frac{\epsilon_{1,j}^{\scriptscriptstyle{(0)}}t}{2}\right)\,\operatorname{Re}\left(\frac{\operatorname{V}_{1,j}^{\scriptscriptstyle{(N)}}}{\epsilon_{j,1}^{\scriptscriptstyle{(0)}}}\overline{\sum_{k\neq j}
		\frac{\operatorname{V}_{1,k}^{\scriptscriptstyle{(N)}}\operatorname{V}_{k,j}^{\scriptscriptstyle{(N)}}}{\epsilon_{j,1}^{\scriptscriptstyle{(0)}}\epsilon_{j,k}^{\scriptscriptstyle{(0)}}}}	\right)
	\nonumber\\
	&-8\,\varepsilon^{4}\sum_{j\neq 1}\sin^{2}\left(\frac{\epsilon_{1,j}^{\scriptscriptstyle{(0)}}t}{2}\right)\,\operatorname{Re}\left(\frac{\operatorname{V}_{1,j}^{\scriptscriptstyle{(N)}}}{\epsilon_{j,1}^{\scriptscriptstyle{(0)}}}\overline{\left(\sum_{k\neq j}
		\sum_{l\neq j}
		\frac{\operatorname{V}_{1,l}^{\scriptscriptstyle{(N)}}\operatorname{V}_{l,k}^{\scriptscriptstyle{(N)}}\operatorname{V}_{k,j}^{\scriptscriptstyle{(N)}}}{\epsilon_{j,1}^{\scriptscriptstyle{(0)}}\epsilon_{j,l}^{\scriptscriptstyle{(0)}}\epsilon_{j,k}^{\scriptscriptstyle{(0)}}}
		-\epsilon_{j}^{\scriptscriptstyle{(2)}}
		\frac{\operatorname{V}_{1,j}^{\scriptscriptstyle{(N)}}}{\epsilon_{j,1}^{\scriptscriptstyle{(0)}}{}^{2}}\right)
	}
	\right)
	\nonumber\\&	-4\,\varepsilon^{4}\sum_{j\neq 1}\left(\sin\left (\epsilon_{1,j}^{\scriptscriptstyle{(0)}}\,t\right )\sin\left(\frac{\varepsilon^{2}\epsilon_{1,j}^{\scriptscriptstyle{(2)}}\,t}{2}\right)\,\frac{|\operatorname{V}_{1,j}^{\scriptscriptstyle{(N)}}|^{2}}{\epsilon_{1,j}^{\scriptscriptstyle{(0)}}{}^{2}}+\sin^{2}\left(\frac{\epsilon_{1,j}^{\scriptscriptstyle{(0)}}t}{2}\right)\,\left |\sum_{k\neq j}
	\frac{\operatorname{V}_{1,k}^{\scriptscriptstyle{(N)}}\operatorname{V}_{k,j}^{\scriptscriptstyle{(N)}}}{\epsilon_{j,1}^{\scriptscriptstyle{(0)}}\epsilon_{j,k}^{\scriptscriptstyle{(0)}}} \right |^{2}\right)
	\nonumber\\
	&	-4\,\varepsilon^{4}\sum_{i> 1}\sum_{j>i}\sin^{2}\left(\frac{\epsilon_{i,j}^{\scriptscriptstyle{(0)}}\,t}{2}\right)\,
	\,\frac{|\operatorname{V}_{1,i}^{\scriptscriptstyle{(N)}}|^{2}}{\epsilon_{1,i}^{\scriptscriptstyle{(0)}}{}^{2}}\,\frac{|\operatorname{V}_{1,j}^{\scriptscriptstyle{(N)}}|^{2}}{\epsilon_{1,j}^{\scriptscriptstyle{(0)}}{}^{2}}+O(\varepsilon^{5})
	\label{hopt:sp}
\end{align}
where in addition to the definitions (\ref{mqm-pt:sp-pt}) we make use of
\begin{align}
	\epsilon_{i}^{\scriptscriptstyle{(2)}}=\sum_{j\neq i}\frac{\left|\operatorname{V}_{i,j}^{\scriptscriptstyle{(N)}}\right |^{2}}{\epsilon_{i,j}^{\scriptscriptstyle{(0)}}}
\end{align}
to denote the leading order correction to the energy levels. 

We can also recover the same result (\ref{hopt:sp}) using Rayleigh-Schr\"odinger time independent perturbation theory  \cite[\S~10.2]{GaPa1991}. 
In that case, we compute the eigenvalues and eigenvectors up to the desired accuracy, and then substitute the result in the general expression (\ref{sp:sp}) of the survival probability for a quantum system with pure point spectrum.

\section{Survival amplitude of the non-relativistic Lee model}
\label{app:Lee}

In section~\ref{sec:mqm-Lee} we prove that the survival probability amplitude of a central qubit in an infinite environment equals that of a non-relativistic Lee model
\begin{align}
&	\alpha(t)=\int_{\mathbb{R}+\imath\varepsilon}\frac{\mathrm{d}z}{2\,\pi\,\imath}\frac{e^{-\imath\,z\,t}}{z-\omega+\omega\,\varkappa^{2}\int_{-\Delta}^{\Delta}\mathrm{d}u \frac{1}{u+\omega-z}}  && \operatorname{Im}z>0
\end{align}

\begin{description}[style=unboxed,leftmargin=0cm]
	\item[The Laplace transform] 
	\begin{align}
		\hat{\alpha}^{\scriptscriptstyle{(up)}}=\frac{1}{z-\omega+\omega\,\varkappa^{2}\int_{-\Delta}^{\Delta}\mathrm{d}u \frac{1}{u+\omega-z}}
	\end{align}
	of the probability amplitude by construction is only defined on the upper half plane of $\mathbb{C}$ where it must be analytic. This is a consequential requirement on the level shift function:
	\begin{align}
		\tilde{\Lambda}^{\scriptscriptstyle{(up)}}(z)=\int_{-\Delta}^{\Delta}\mathrm{d}u \frac{1}{u+\omega-z}
	\end{align}
	Whereas the evaluation of the real part is straightforward
	\begin{align}
		\operatorname{Re}\tilde{\Lambda}^{\scriptscriptstyle{(up)}}(x+\imath\,y)=\int_{-\Delta}^{\Delta}\mathrm{d}u \frac{u+\omega-x}{(u+\omega-x)^{2}+y^{2}}=\frac{1}{2}\ln \frac{(\omega+\Delta-x)^{2}+y^{2}}{(\omega-\Delta-x)^{2}+y^{2}}
	\end{align}
	the imaginary part 
	\begin{align}
		\operatorname{Im}\tilde{\Lambda}^{\scriptscriptstyle{(up)}}(x+\imath\,y)
		=\int_{\frac{y}{\omega+\Delta-x}}^{\frac{y}{\omega-\Delta-x}}\mathrm{d}u\frac{1}{1+u^{2}}
		\nonumber
	\end{align}
	requires extra care.	Namely, the boundary of integration diverge at the points
	\begin{align}
		x=\{ \omega-\Delta, \omega+\Delta\}
	\end{align} 
	For any positive $y$, we obtain a continuous dependence on $x$ by setting
	\begin{align}
		\operatorname{Im}\tilde{\Lambda}^{\scriptscriptstyle{(up)}}(x+\imath\,y)&=\pi\,\Big{(}\theta(\omega+\Delta-x)-\theta(\omega-\Delta-x)\Big{)}
		\nonumber\\
&		+\arctan\frac{y}{\omega+\Delta-x}-\arctan\frac{y}{\omega-\Delta-x}
	\end{align}
	where $\theta$ is the Heaviside step function. 	The result is exactly what we get on the upper half-plane 
	by applying the formal definition of the logarithm of a complex number (\ref{mqm-Lee:log}) used in the main text.
	\item[The analytic extension] of the Laplace transform to the lower half plane allows us to apply the Cauchy theorem and other tools of complex analysis to evaluate the survival amplitude. We define 
	\begin{align}
		&	\hat{\alpha}^{\scriptscriptstyle{(lo)}}(x-\imath\,y):=\overline{\hat{\alpha}^{\scriptscriptstyle{(up)}}(x+\imath\,y)} &&\forall\,x\,\in\,\mathbb{R} & \&&&\forall\,y\,>\,0
		\nonumber
	\end{align}
	which implies that for $y<0$
	\begin{align}
	\operatorname{Im}\tilde{\Lambda}^{\scriptscriptstyle{(lo)}}(x+\imath\,y)&=-\,\pi\,\Big{(}\theta(\omega+\Delta-x)-\theta(\omega-\Delta-x)\Big{)}
	\nonumber\\
	&		+\arctan\frac{y}{\omega+\Delta-x}-\arctan\frac{y}{\omega-\Delta-x}
	\end{align}
	Hence, we are in a position to extend the Laplace transform of the survival probability to a function on the full complex plane
	\begin{align}
		\hat{\alpha}(z)=\begin{cases}
			\hat{\alpha}^{\scriptscriptstyle{(up)}}(z)	& \operatorname{Im}z\,\geq\,0
			\\[0.3cm]
			\hat{\alpha}^{\scriptscriptstyle{(lo)}}(z)	& \operatorname{Im}z\,<\,0
		\end{cases}
	\end{align}
	By construction, the extension is analytic outside the real axis, in agreement with the axiomatic requirements imposed on the resolvent by unitary of a quantum evolution \cite[\S~A.III.3]{CoTaDuRoGr1998}. The analytic extension is discontinuous across a cut for 
	\begin{align}
		z\in [\omega-\Delta,\omega+\Delta]
		\label{cut}
	\end{align} 
	where we find
	\begin{align}
		\lim_{y\downarrow 0}\Big{(}\hat{\alpha}(x+\imath\,y)-\hat{\alpha}(x-\imath\,y)\Big{)}=-2\,\imath\,\lim_{y\downarrow 0}\,|\hat{\alpha}(x+\imath\,y)|^{2}\tilde{\Lambda}^{\scriptscriptstyle{(up)}}(x+\imath\,y)
	\end{align} 
	\item[The extension to the second Riemann sheet] of the Laplace transform follows from requiring continuity along contours crossing the cut:
	\begin{align}
		\hat{\alpha}_{\scriptscriptstyle{II}}(z)=\begin{cases}
			\hat{\alpha}^{\scriptscriptstyle{(lo)}}(z)	& \operatorname{Im}z\,\geq\,0
			\\[0.3cm]
			\hat{\alpha}^{\scriptscriptstyle{(up)}}(z)	& \operatorname{Im}z\,<\,0
		\end{cases}
		\label{ac}
	\end{align} 
	This means extending the functional form of $\hat{\alpha}^{(up)}$, $\hat{\alpha}^{(lo)}$ to arguments outside the original domain of definition. 
	The definition (\ref{ac}) also implies that 
	\begin{align}
		&	 	\hat{\alpha}_{\scriptscriptstyle{II}}(x)=\alpha(x) && \forall\,x\,\in \mathbb{R}\setminus I
	\end{align}
	where $I$ is the set of the real axis that includes the cut and all the poles of the Laplace transform.
\end{description}

\subsection{Poles on the second Riemann sheet} 
\label{app:Lee-poles}
Dealing with the non-relativistic Lee model offers the advantage that the level shift function and its analytic extension to the second Riemann shift are amenable to exact explicit expressions. We can thus directly verify the presence of poles outside the real axis on the second Riemann sheet. It is, however, instructive to prove this fact in a model independent way. The crucial hypothesis is that the Laplace transform takes the same value on the first and second Riemann sheets outside the cut. This hypothesis  holds true for the non relativistic Lee model but not for the Rosenzweig-Porter model we consider in the main text.   The proof, inspired by \cite{ArOs1963}, is by contradiction: let us assume that $\hat{\alpha}_{\scriptscriptstyle{II}}(z) $ is analytic outside the real axis. The application of the dispersion relations on the second Riemann sheet yields for every $z$ outside $I$ 
\begin{align}
	\hat{\alpha}_{\scriptscriptstyle{II}}(z)&=\frac{1}{\imath\pi}\int_{\omega-\Delta}^{\omega+\Delta}\mathrm{d}x\frac{\hat{\alpha}_{\scriptscriptstyle{II}}(x+\imath\,\varepsilon)-\hat{\alpha}_{\scriptscriptstyle{II}}(x-\imath\varepsilon)}{x-z}+\sum_{x_{i}\,\in\,\mathscr{R}}\frac{\operatorname{Res}[\hat{\alpha}_{\scriptscriptstyle{II}}](x_{i})}{x_{i}-x}
\end{align} 
where $\mathscr{R}$ is the set of the (simple) poles in $I$. The definition of $\hat{\alpha}_{\mathcal{II}} $ also implies
\begin{align}
	\hat{\alpha}_{\scriptscriptstyle{II}}(z)
	&=-\frac{1}{\imath\pi}\int_{\omega-\Delta}^{\omega+\Delta}\mathrm{d}x\frac{\hat{\alpha}(x+\imath\,\varepsilon)-\hat{\alpha}(x-\imath\varepsilon)}{x-z}+\sum_{x_{i}\,\in\,\mathscr{R}}\frac{\operatorname{Res}[\hat{\alpha}](x_{i})}{x_{i}-x}
	\nonumber\\
	&	 	=-\hat{\alpha}(z)+2\sum_{x_{i}\,\in\,\mathscr{R}}\frac{\operatorname{Res}[\hat{\alpha}_{\scriptscriptstyle{II}}](x_{i})}{x_{i}-x}
\end{align}
If we now evaluate this last identity for any $x$ on the real axis outside $I$ we obtain
\begin{align}
	\alpha(x)=\sum_{x_{i}\,\in\,\mathscr{R}}\frac{\operatorname{Res}[\hat{\alpha}_{\scriptscriptstyle{II}}](x_{i})}{x_{i}-x}
	\nonumber
\end{align}
which is satisfied only if
\begin{align}
	\frac{1}{\imath\pi}\int\limits_{\omega-\Delta}^{\omega+\Delta}\mathrm{d}x\frac{\hat{\alpha}(x+\imath\,\varepsilon)-\hat{\alpha}(x-\imath\varepsilon)}{x-z}=0
\end{align}
The latter condition contradicts the existence of a cut.

\subsection{Integration contours on the complex plane}

The analytic extension of the Laplace transform of the survival amplitude to the lower half plane of the complex plane (first Riemann sheet) and subsequently to the second Riemann sheet  opens the way to evaluate the survival probability in three different ways; 
\begin{enumerate}[label=M-\roman*]
	\item \label{M1}by directly performing the integral
	\begin{align}
		\alpha(t)=\int_{\mathbb{R}+\imath\varepsilon}\frac{\mathrm{d}z}{2\,\pi\,\imath}\frac{e^{-\imath\,z\,t}}{z-\omega+\omega\,\varkappa^{2}\,\tilde{\Lambda}^{\scriptscriptstyle{(up)}}(z)}
	\end{align}
or by means of Cauchy theorem by interpreting the inverse Laplace transform as contributing
	\item  \label{M2} to the line-integral along the contour $\mathcal{C}$ of Fig~\ref{fig:contour1} which leads to the identity
	\begin{align}
		\alpha(t)&	=\sum_{x_{i}\in \mathcal{R}}e^{-\imath\,x_{i}\,t}\operatorname{Res}[S](x_{i})
		\nonumber\\
&	+
		\int\limits_{\omega-\Delta}^{\omega+\Delta}\frac{\mathrm{d}x}{2\,\pi\,\imath}\frac{\omega\,\varkappa^{2}\,e^{-\imath\,x\,t}}{\left (x-\omega+\frac{\omega\,\varkappa^{2}}{2}\ln\frac{(x-\omega-\Delta)^{2}}{(x-\omega+\Delta)^{2}}\right )^{2}+\omega^{2}\varkappa^{4}\,\pi^{2}}
		\label{sp-Lee:Riemann1}
	\end{align}
	where $\mathcal{R}^{\prime}$ is the set of poles on the real axis;
	\item \label{M3} or to the line integral along the contour $\mathcal{C}$ of Fig~\ref{fig:contour2}, which gives
	\begin{align}
		\alpha(t)=	\sum_{x_{i}\in \mathcal{R}}e^{-\imath\,x_{i}\,t}\operatorname{Res}[S](x_{i})+
		\sum_{z_{i}\in \mathcal{R}^{\prime}}e^{-\imath\,z_{i}\,t}\operatorname{Res}[S](z_{i})+I^{\scriptscriptstyle{(1)}}(t)-I^{\scriptscriptstyle{(2)}}(t)
		\label{sp-Lee:Riemann2}
	\end{align}
	where $\mathcal{R}^{\prime}$ is the set of poles on the second Riemann sheet and
	\begin{align}
	&	I^{\scriptscriptstyle{(1)}}(t)=\int\limits_{\omega-\Delta-\imath\,\infty}^{\omega-\Delta}\frac{\mathrm{d}z}{2\,\pi\,\imath}
		\left(
		\frac{e^{-\imath\,u\,t}}{u-\omega+\omega\,\varkappa^{2}\tilde{\Lambda}^{\scriptscriptstyle{(lo)}}(u)}
		-
		\frac{e^{-\imath\,u\,t}}{u-\omega+\omega\,\varkappa^{2}\tilde{\Lambda}^{\scriptscriptstyle{(lo)}}(u)-2\,\imath\,\varkappa^{2}\,\pi}
		\right)
		\\
	&	I^{\scriptscriptstyle{(2)}}(t)=\int\limits_{\omega+\Delta-\imath\,\infty}^{\omega+\Delta}\frac{\mathrm{d}z}{2\,\pi\,\imath}
		\left(
		\frac{e^{-\imath\,(u+\omega+\Delta)\,t}}{u-\omega+\omega\,\varkappa^{2}\tilde{\Lambda}^{\scriptscriptstyle{(lo)}}(u)}
		-
		\frac{e^{-\imath\,(u+\omega+\Delta)\,t}}{u-\omega+\omega\,\varkappa^{2}\tilde{\Lambda}^{\scriptscriptstyle{(lo)}}(u)-2\,\imath\,\omega\,\varkappa^{2}\,\pi}
		\right )
	\end{align}
\end{enumerate}
The method \ref{M2} offers a good numerical alternative to \ref{M1}, especially at weak coupling.  Namely, the detailed analysis carried out in \cite{WolT2013}  proves that at weak coupling the residues of the poles  on the real axis 
vanish exponentially fast. As the coupling strength increases,  \ref{M2} the contribution of  the residues of the poles on the real axis becomes more important and, ultimately, dominates the dynamics. The method \ref{M3} explicitly extricates the intermediate asymptotic exponential associated to the
pole on the second Riemann sheet closest to the real axis.

\section{Proof of the Kroenecker-Weyl's phase averaging theorem}
\label{app:KW}

Suppose that $f$ is a multivariate function, $2\,\pi$ periodic with respect to all its, say $N$, arguments.  We are interested in the time average
\begin{align}
	A=\lim_{T\uparrow \infty}\frac{1}{ \,T}\int_{-T}^{T}\mathrm{d}t\,f(\phi_{\mathfrak{1}}(t),\dots,\phi_{N}(t))
\end{align}
along trajectories that depend linearly on the time variable:
\begin{align}
&	\phi_{\ell}(t)=\epsilon_{\ell}\,t+\alpha_{\ell}
&& \ell=1,\dots,N
\end{align}
To this end, we resort to the Fourier series representation of the function
\begin{align}
	f(x_{\mathfrak{1}},\dots,x_{N})=
	\sum_{n_{\mathfrak{1}},\dots,n_{N}=-\infty}^{\infty}e^{\imath\,2\pi\sum_{\ell=1}^{N}n_{\ell}\,x_{\ell}}\,\hat{f}_{n_{\mathfrak{1}},\dots,n_{N}}
\end{align}
and insert it into the time average. We are thus in a position to evaluate the time integral. The result is
\begin{align}
A		=\sum_{n_{\mathfrak{1}},\dots,n_{N}=-\infty}^{\infty}\hat{f}_{n_{\mathfrak{1}},\dots,n_{N}}\,\lim_{T\uparrow \infty}\,\frac{1}{T}\prod_{i=1}^{n} \frac{\sin(\sum_{\ell=1}^{N}\,n_{\ell}\,\epsilon_{\ell}\,T)}{\sum_{\ell=1}^{N}n_{\ell}\,\epsilon_{\ell}} e^{\imath\,2\,\pi\,n_{\ell}\alpha_{\ell}}
\end{align}
If we now assume that all the frequencies are incommensurate
\begin{align}
&	\sum_{\ell=1}^{N}n_{\ell}\,\epsilon_{\ell}=0
&& \Leftrightarrow & n_{\ell}=0&& \forall\,\ell=1,\dots,N
\end{align}
the limit suppresses all terms of the series except the first
\begin{align}
	A	=
	\hat{f}_{0,\dots,0}=\prod_{\ell=1}^{N}\int_{0}^{2\pi}\frac{\mathrm{d}x_{\ell}}{2\,\pi}\,	f(x_{\mathfrak{1}},\dots,x_{N})
\end{align}
which yields the claim.

\section{Integrals evaluated to obtain the  extension of the Kac-Mazur-Montroll estimate}
\label{app:Integrals}

\subsection{Evaluation of the Kroenecker-Weyl phase averaging theorem}

We want to derive (\ref{Kac:pares1}), (\ref{Kac:pares2}) from (\ref{Kac:start}).
The apparent stumbling block is that the functions $V(\phi)$ and $W(\phi)$ are quadratic and not linear in the trigonometric functions.  Indeed, if we define
\begin{align}
&	\begin{cases}
	\alpha_{1}:=\sum_{\ell=1}^{N} \cos \phi_{\ell}\,c_{\scriptscriptstyle{\ell}}
\\[0.3cm]
 \alpha_{2}:=\sum_{\ell=1}^{N} \sin \phi_{\ell}\,c_{\scriptscriptstyle{\ell}}
	\end{cases}
	&& \& &
\begin{cases}
	\beta_{1}=2\sum_{k=1}^{N}\epsilon_{k}\sin \phi_{k}\,c_{\scriptscriptstyle{k}}
	\\[0.3cm]
	\beta_{2}=2\sum_{k=1}^{N}\epsilon_{k}\cos \phi_{k}\,c_{\scriptscriptstyle{k}}
\end{cases}
\end{align}
we can write $V(\phi)$ and $W(\phi)$ in (\ref{Kac:start}) as:
\begin{align}
	&	V(\bm{\phi})
	=\alpha_{1}^{2}+\alpha_{2}^{2}
\\
	&W(\bm{\phi})
	=-\alpha_{1}\beta_{1}+\alpha_{2}\beta_{2}
\end{align}
To turn the quadratic dependence into a linear one, we resort to the complex  Hubbard-Stratonovich transformation:
\begin{align}
	&	e^{\imath\,z\,x^{2}}=\frac{2\sqrt{z}}{\sqrt{\pi}}e^{\imath\frac{\pi}{4}}\int_{\mathbb{R}}\mathrm{d}a\,e^{-4\,\imath\,z\,a^{2}+4\,\imath \,z\,x\,a} && z>0
\\
	&	e^{-\imath\,z\,x^{2}}=\frac{2\sqrt{z}}{\sqrt{\pi}}e^{-\imath\frac{\pi}{4}}\int_{\mathbb{R}}\mathrm{d}a\,e^{4\,\imath\,z\,a^{2}-4\,\imath \,z\,x\,a} && z>0
\end{align}
For perfect squares we get
\begin{align}
	e^{\imath\,z \,(\alpha_{1}^{2}+\alpha_{2}^{2})}
	=\frac{4\,z\,e^{\imath\frac{\pi}{2}}}{\pi}\int_{\mathbb{R}^{2}}\mathrm{d}a_{1}\mathrm{d}a_{2}
	e^{-4\,\imath\,z\,(a_{1}^{2}+a_{2}^{2})+4\,\imath \,z\,(\alpha_{1}\,a_{1}+\alpha_{2}\,a_{2})} 
\end{align}
whereas for cross-products we obtain the identities
\begin{align}
	&	e^{-\imath \,y\alpha_{1}\,\beta_{1}}=e^{-\imath \,y\frac{(\alpha_{1}+\beta_{1})^{2}}{4}}e^{\imath\,y\, \,\frac{(\alpha_{1}-\beta_{1})^{2}}{4}}=\frac{y}{\pi}\int_{\mathbb{R}^{2}}\prod_{i=1}^{2}\mathrm{d}b_{i}
	e^{\imath\,y\,(b_{1}^{2}-b_{2}^{2})-\imath \,y\,((\alpha_{1}+\beta_{1})\,b_{1}-(\alpha_{1}-\beta_{1})\,\,b_{2})} 
	\nonumber\\
	&	e^{\imath \,y\alpha_{2}\,\beta_{2}}=e^{\imath \,y\frac{(\alpha_{2}+\beta_{2})^{2}}{4}}e^{-\imath\,y\, \,\frac{(\alpha_{2}-\beta_{2})^{2}}{4}}=\frac{y}{\pi}\int_{\mathbb{R}^{2}}\prod_{i=4}^{4}\mathrm{d}b_{i}
	e^{-\imath\,y\,(b_{3}^{2}-b_{4}^{2})+\imath \,y\,((\alpha_{2}+\beta_{2})\,b_{3}-(\alpha_{2}-\beta_{2})\,\,b_{4})} 
\end{align}
The upshot is that we need to evaluate
\begin{align}
	\nu(b)=8\,\operatorname{Re}\imath \int_{\mathbb{R}}\frac{\mathrm{d}a_{1}\mathrm{d}a_{2}}{\pi}\int_{\mathbb{R}}\frac{\prod_{i=1}^{4}\mathrm{d}b_{i}}{\pi^{2}}
	\int_{\mathbb{R}_{+}^{2}}\frac{\mathrm{d}z\mathrm{d}y}{2\,\pi^{2}}
	e^{-4\,\imath\,z\,(a_{1}^{2}+a_{2}^{2})}\,	e^{-\imath\,y\, (b_{1}^{2}-b_{2}^{2}-b_{3}^{2}-b_{2}^{2}+b_{4}^{2})}\,e^{-\imath\,p\,z}\, z\,
	\tilde{I}(z,y,\bm{a},\bm{b}) 	
\end{align}
where
\begin{align}
	\tilde{I}(z,y,\bm{a},\bm{b})=\int_{0}^{2\pi}\prod_{i=1}^{N}\frac{\mathrm{d}\phi_{i}}{2\,\pi}\,
	e^{4\,\imath \,z\,(\alpha_{1}\,a_{1}+\alpha_{2}\,a_{2})} 
	\left(1-e^{\imath \,y\,\Big{(}-(\alpha_{1}+\beta_{1})\,b_{1}+(\alpha_{1}-\beta_{1})\,b_{2}+(\alpha_{2}+\beta_{2})\,b_{3}-(\alpha_{2}-\beta_{2})\,b_{4}\Big{)}} \right)
\end{align}
We are now in a position to use the identity (\ref{Kac:Bessel}) to evaluate the time average.  Namely, we first use the identities
\begin{align}
	&	4 \,z\,(\alpha_{1}\,a_{1}+\alpha_{2}\,a_{2})+ \,y\,\Big{(}-(\alpha_{1}+\beta_{1})\,b_{1}+(\alpha_{1}-\beta_{1})\,b_{2}+(\alpha_{2}+\beta_{2})\,b_{3}-(\alpha_{2}-\beta_{2})\,b_{4}\Big{)} 
	\nonumber\\
	&=\alpha_{1}\left(4 \,z\,a_{1}-y\,(b_{1}-b_{2})\right)+\alpha_{2}\left(4 \,z\,a_{2}+y\,(b_{3}-b_{4})\right)
	-y\,\beta_{1}\,\left(b_{1}+b_{2}\right)+y\,\beta_{2}\,\left(b_{3}+b_{4}\right)
	\nonumber\\
	&=\sum_{k=1}^{N}c_{k}\,\left(\cos\phi_{k}\Big{(}4 \,z\,a_{1}-y\,(b_{1}-b_{2})+2\,\epsilon_{k}\,y\,\left(b_{3}+b_{4}\right)\Big{)}
	+\sin\phi_{k}\Big{(}4 \,z\,a_{2}+y\,(b_{3}-b_{4})-2\,\epsilon_{k}\,y\,\left(b_{1}+b_{2}\right)\Big{)}\right)
\end{align}
Next, we apply (\ref{Kac:Bessel}) to evaluate the integrals over the phase-variables. We obtain:
\begin{align}
&	\tilde{I}(z,y,\bm{a},\bm{b}) 	=\prod_{j=1}^{N} \mathscr{J}_{0}\left(4\,c_{\scriptscriptstyle{j}}\, z\sqrt{a_{\mathfrak{1}}^{2}+a_{\mathfrak{2}}^{2}}\right)
	\nonumber\\
	&-\prod_{j=1}^{N}\mathscr{J}_{0}\left(c_{\scriptscriptstyle{j}} \sqrt{\big{(}4 \,z\,a_{1}-y\,(b_{1}-b_{2})+2\,\epsilon_{k}\,y\,\left(b_{3}+b_{4}\right)\big{)}^{2}+\big{(}4 \,z\,a_{2}+y\,(b_{3}-b_{4})-2\,\epsilon_{k}\,y\,\left(b_{1}+b_{2}\right)\big{)}^{2}}\right)
\end{align}
The last step is to perform the change of variables
\begin{align}
	&	\begin{cases}
		x_{1}=b_{1}-b_{2}
		\\[0.3cm]
		x_{2}=b_{1}+b_{2}
	\end{cases}
	&&	
	\begin{cases}
		x_{3}=-b_{3}+b_{4}
		\\[0.3cm]
		x_{4}=b_{3}+b_{4}
	\end{cases}
\end{align}
We get
\begin{align}
	&	\nu(b)=2\,\operatorname{Re}\imath\int_{\mathbb{R}_{+}^{2}}\frac{\mathrm{d}z\mathrm{d}y}{2\,\pi^{2}}
	\,e^{-\imath\,p\,z}\,z \int_{\mathbb{R}^{2}}\frac{\mathrm{d}a_{1}\mathrm{d}a_{2}}{\pi}\int_{\mathbb{R}^{4}}\frac{\prod_{i=1}^{4}\mathrm{d}x_{i}}{\pi^{2}}
	e^{-4\,\imath\,z\,(a_{1}^{2}+a_{2}^{2})}\,e^{-\imath\,y\, (x_{1}x_{2}+x_{3}x_{4})}
	\nonumber\\
	&	\,\times\,\prod_{j=1}^{N} \left(\mathscr{J}_{0}\left(4\,c_{\scriptscriptstyle{j}}\, z\sqrt{a_{\mathfrak{1}}^{2}+a_{\mathfrak{2}}^{2}}\right)-\mathscr{J}_{0}\left(c_{\scriptscriptstyle{j}} \sqrt{\big{(}4 \,z\,a_{1}-y\,x_{1}+2\,\epsilon_{k}\,y\,x_{4}\big{)}^{2}+\big{(}4 \,z\,a_{2}-y\,x_{3}-2\,\epsilon_{k}\,y\,x_{2}\big{)}^{2}}\right)\right)
\end{align}

\subsection{Integration over the Hubbard-Stratonovich auxiliary variables}

The integrals over the auxiliary variables $x_{1},x_{2},x_{3},x_{4}$ are Gaussian. The result of these integrals can be checked with any software for symbolic manipulation. We arrive at
\begin{align}
	\nu(p)=4\,\operatorname{Re}\imath \int_{\mathbb{R}_{+}^{2}}\frac{\mathrm{d}z\mathrm{d}y}{2\,\pi^{2}}
	\int_{\mathbb{R}^{2}}\frac{\mathrm{d}a_{1}\mathrm{d}a_{2}}{\pi}e^{-4\,\imath\,z\,(a_{1}^{2}+a_{2}^{2})} \,e^{-\imath\,p\,z}\,z\left(\frac{e^{-4\,\varkappa\,z^{2}\,(a_{\mathfrak{1}}^{2}+a_{\mathfrak{2}}^{2})}}{y^{2}}
	-\frac{e^{-\frac{4\,\varkappa\,z^{2}\,(a_{\mathfrak{1}}^{2}+a_{\mathfrak{2}}^{2})}{\,(1+y^{2}\,(\varkappa\,\Gamma-\gamma^{2})}}}{y^{2}\,(1+y^{2}\,(\varkappa\,\Gamma-\gamma^{2})}\right)
\end{align}
This integral can be performed in polar variables. The angular part yields
\begin{align}
	\nu(p)=4\,\operatorname{Re}\imath \int_{\mathbb{R}_{+}^{2}}\frac{\mathrm{d}z\mathrm{d}y}{\pi^{2}}\, z
	\int_{0}^{\infty}\mathrm{d}a \,a e^{-4\,\imath\,z\,a^{2}-\imath\,z\,p} \left(e^{-4\,\varkappa\,z^{2}\,a^{2}}
	-\frac{e^{-\frac{4\,\varkappa\,z^{2}\,a^{2}}{1+y^{2}\,(\varkappa\,\Gamma-\gamma^{2})}}}{y^{2}\,(1+y^{2}\,(\varkappa\,\Gamma-\gamma^{2})}\right)
\end{align}
The radial integral yields 
\begin{align}
	\nu(p)=\frac{1}{2\,\pi^{2}}\,\operatorname{Re}\imath \int_{\mathbb{R}_{+}^{2}}\mathrm{d}z\mathrm{d}y\,
	\frac{e^{-\imath\,z\,p}}{y^{2}} \left(\frac{1}{\imath+\varkappa\,z}
	-\frac{1}{(1+y^{2}\,(\varkappa\,\Gamma-\gamma^{2}))\left(\imath+\frac{\varkappa\,z}{1+y^{2}\,(\varkappa\,\Gamma-\gamma^{2})}\right)}\right)
\end{align}
which is the expression in the main text.

\bibliography{revival}{} 


\end{document}